\documentclass{JHEP3}

  \usepackage{latexsym,bm,amsmath,amssymb,amsfonts}
  \usepackage{epsfig,graphics,graphicx}
  \usepackage{slashed}

  \long\def\comment#1{ }

  \newcommand{\dif}{{\rm d}}

  \newcommand{\del}{\partial}

  \newcommand{\rme}{{\rm e}}
  \newcommand{\rmd}{{\rm d}}   %ELS%

  \newcommand{\nn}{\nonumber\\}

  \newcommand{\beq}{\begin{eqnarray}}
  \newcommand{\eeq}{\end{eqnarray}}
  
 \def\simge{\mathrel{%
   \rlap{\raise 0.511ex \hbox{$>$}}{\lower 0.511ex \hbox{$\sim$}}}}
\def\simle{\mathrel{
   \rlap{\raise 0.511ex \hbox{$<$}}{\lower 0.511ex \hbox{$\sim$}}}}

\title{\rm \LARGE Jet evolution in the ${\mathcal N}=4$
SYM plasma at strong coupling}

\author{Y.~Hatta\\Graduate School of Pure and Applied Sciences, University
of Tsukuba, Tsukuba, Ibaraki 305-8571, Japan\\
E-mail: \email
{hatta@het.ph.tsukuba.ac.jp
 }}
\author{E. Iancu\\Institut de Physique Th\'eorique de Saclay,
 F-91191 Gif-sur-Yvette, France\\
        E-mail: \email{Edmond.Iancu@cea.fr}}
\author{A. H.~Mueller\\Department of Physics, Columbia University, New York, NY
10027, USA\\
        E-mail: \email{amh@phys.columbia.edu}}

\abstract{
Within the framework of the AdS/CFT correspondence, we study the time
evolution of an energetic ${\mathcal R}$--current propagating through
a finite temperature, strongly coupled, ${\mathcal N}=4$ SYM plasma
and propose a physical picture for our results. In this picture,
the current splits into a pair of massless partons, which then evolve
via successive branchings, in such a way that energy is
quasi--democratically divided among the products of a branching.
We point out a duality between the transverse size of the partonic system
produced through branching and the radial distance traveled by the dual
Maxwell wave in the AdS geometry. For a time--like current, the branching
occurs already in the vacuum, where it gives rise to a system of
low--momentum partons isotropically distributed in the transverse plane.
But at finite temperature, the branching mechanism is modified by the
medium, in that the rate for parton splitting is enhanced
by the transfer of transverse momentum from the partons to the plasma.
This mechanism, which controls the parton energy loss, is sensitive to the
energy density in the plasma, but not to the details of the thermal state.
We compute the lifetime of the current for various
kinematical regimes and provide physical interpretations for other,
related, quantities, so like the meson screening length, the drag force,
or the trailing string, that were previously computed via AdS/CFT techniques.
}

\begin{document}

\section{Introduction}
\label{Intro}

Motivated by some experimental results from the heavy ion program at
RHIC, which suggest that the deconfined, `quark--gluon', matter produced
in the early stages of an ultrarelativistic nucleus--nucleus collision
might be strongly interacting (see, e.g., the review papers
\cite{Gyulassy:2004zy,Muller:2007rs} and references therein), there was
recently an abondance of applications of the AdS/CFT correspondence to
problems involving a strongly--coupled gauge plasma at finite temperature
and/or finite quark density (for a recent review see \cite{Son:2007vk}).
While the early such applications have focused on the long--range and
large--time properties of the plasma, so like hydrodynamical flow and
transport coefficients, more recent studies have been also concerned with
the response of the plasma to a `hard probe' --- an energetic `quark' or
`current' which probes the plasma on very short time and distance scales,
much shorter than the thermal scale $1/T$ (with $T$ being the
temperature). Although the relevance of such applications to actual hard
probes in QCD is perhaps not so clear (since, by asymptotic freedom, QCD
should be weakly coupled on such short space--time separations), the
results that have been obtained in this way are conceptually interesting,
in that they shed light on a new physical regime --- that of a gauge
theory with strong interactions --- which for long time precluded all
first--principles theoretical investigations other than lattice gauge
theory. For these results to be accompanied by conceptual clarifications,
a physical interpretation for them is strongly needed, but this seems to
be difficult without direct calculations in the original gauge theory. A
different strategy, which is less rigorous, is to propose a physical
picture based on general arguments and then demonstrate that this picture
is consistent with all the available results (to the extent that
comparisons are possible). This will be our strategy in this paper.

The physical problem that we shall consider --- the propagation of an
`electromagnetic' current (actually, an ${\mathcal R}$--current) through
the plasma --- is particularly well suited for our purposes since, first,
it has a strong overlap with several other problems previously considered
in the literature and, second, it does not require any extension of the
AdS/CFT correspondence (like the introduction of additional D7--branes)
beyond the `minimal' framework of its original formulation
\cite{Maldacena:1997re,Gubser:1998bc,Witten:1998zw} --- so it avoids any
potential artifact due to such extensions. Besides, this problem has
another useful feature: for the gauge theory of interest
--- namely, the conformally invariant ${\mathcal N}=4$ supersymmetric
Yang--Mills (SYM) theory --- the current--current correlator in the
vacuum (or `vacuum polarization tensor') is protected by supersymmetry
\cite{Anselmi:1997am}, in such a way that the full result in the strong
`t Hooft coupling limit $\lambda\equiv g^2N_c\to\infty$ is exactly the
same as the corresponding result in lowest--order perturbation theory
(i.e., the one--loop approximation). This property will allow us to
recognize --- via a comparison between the space--time picture of the
one--loop process in the gauge theory and the `supergravity' picture of
the dual AdS/CFT calculation --- an interesting `duality' between the
physical transverse size of the partonic system into which the current is
evolving and the radial distance in $AdS_5$. In turn, this duality will
be a key ingredient of our proposal for a physical interpretation.

%Of course, the actual physical picture behind this result is very
%different at weak and, respectively, strong coupling, but the one--loop
%picture must be {\em formally} consistent with the AdS/CFT calculation at
%strong coupling, as it provides the same final result. This simple observation

Specifically, we shall follow the time--evolution of an ${\mathcal
R}$--current which in the plasma rest frame propagates like a plane--wave
in the $z$ direction, with a large longitudinal momentum $q\gg T$ and
frequency $\omega \simeq q$, and which at $t=0$ has zero transverse size.
At weak coupling, this current would develop a partonic fluctuation
involving just two partons --- massless fields from ${\mathcal N}=4$ SYM
which carry ${\mathcal R}$--charge and transform in the adjoint
representation of the color group SU$(N_c)$. For a space--like current,
this pair would grow in transverse space up to a maximal size $L\sim 1/Q$
and live for a relatively long time $t_c\sim q/Q^2$, thus acting like a
`color dipole' which can mediate the current interactions with an
external target. (Here $Q^2\equiv q^2-\omega^2$ is the virtuality of the
current, and is positive in the space--like case.) For a time--like
current ($Q^2 < 0$), the pair can dissociate after a time $t\sim t_c$ and
thus give rise to two free partons which separate from each other, so
like the quark and antiquark jets produced in $e^+e^-$ annihilation. This
perturbative, one--loop, picture formally applies also to the full vacuum
polarization tensor at strong coupling, because of the
non--renormalization property alluded to above. This explains the
polyvalence of the ${\mathcal R}$--current as a `hard probe', including
at strong coupling\footnote{We should also mention here the use of the
current in the calculation of the rate for `photons' or `dileptons'
production in the ${\mathcal N}=4$ SYM plasma at strong coupling
\cite{CaronHuot:2006te}.} : by tuning the virtuality and the momentum of
the current, we can mimic a color dipole, a `meson', or a pair of jets
with the desired values for the system size and rapidity, and then study
the interactions between this partonic system and the plasma (or any
other target).

But, of course, all that applies to a current propagating through the
vacuum, and there is {\em a priori} no guarantee that a similar strategy
should also work in the thermal bath. In a previous analysis
\cite{Hatta:2007cs}, we have considered the case of a space--like current
--- i.e., the problem of deep inelastic scattering off the plasma --- and
shown (via the appropriate AdS/CFT calculation) that this strategy still
works, but only for not too high energies. A current with relatively low
energy, such that $q\ll Q^3/T^2$, propagates through the plasma
essentially without interacting, so like a `small meson' in the approach
of Refs.
\cite{Peeters:2006iu,Liu:2006nn,Chernicoff:2006hi,Caceres:2006ta,Avramis:2006em,Liu:2006he},
where the `meson' was made with two `heavy quarks' attached to a
D7--brane. Note that the above condition on $q$ can be rewritten as a
lower bound on the current virtuality, $Q\gg Q_s$, with $Q_s\sim
(qT^2)^{1/3}$ the plasma saturation momentum, and also as an upper bound
on the transverse size of the effective dipole, $L\ll 1/Q_s$, which is
then consistent with the respective bound (the `meson screening length')
found in Refs. \cite{Peeters:2006iu,Liu:2006nn,Chernicoff:2006hi}.

But for higher energies $q\gtrsim Q^3/T^2$, or, equivalently, lower
virtualities $Q\lesssim Q_s$, the analysis of Ref. \cite{Hatta:2007cs}
shows that the current is very rapidly absorbed into the plasma, over a
time $t_s\sim q/Q_s^2(q)\propto q^{1/3}$ which is much shorter than the
period $t_c\sim q/Q^2$ required for the formation of a nearly on--shell
partonic fluctuation. That is, for such a high energy, the current cannot
be assimilated with a color dipole anymore (not even over a finite period
of time), as it disappears before a dipole can form. The above result on
$t_s$ can be restated by saying that the current propagates through the
plasma over a longitudinal distance $z_s \propto q^{1/3}$ before it
disappears. Interestingly, the same parametric estimate has been very
recently found for the penetration length of an effective gluon
\cite{Gubser:2008as}. In view of the physical picture that we shall
develop, this similarity is not just a coincidence, but rather it
reflects the universality of the dissipation mechanism in the
strongly--coupled plasma.

In Ref. \cite{Hatta:2007cs}, the current lifetime $t_s$ has been inferred
from physical considerations, but no temporal evolution (on top of the
usual phase $\rme^{-i\omega t}$ from the definition of the plane--wave)
was explicitly considered. In this paper we shall extend that analysis by
addressing the time--dependent problem, for both space--like and
time--like currents, with the purpose of elucidating the dynamics of the
current and the mechanism responsible for its dissipation. As already
mentioned, we shall consider initial conditions such that the current has
zero transverse size at $t=0$. In the dual string calculation, this is
represented by a vector field propagating in the $AdS_5$--Schwarzschild
metric which at $t=0$ is localized at the Minkowski boundary $r\to\infty$
of $AdS_5$. For $t>0$, this perturbation propagates inside the bulk of
the `5th dimension' as a Maxwell wave, i.e., according to the Maxwell
equations in the $AdS_5$--Schwarzschild geometry. For the kinematics of
interest, these equations can be suggestively rewritten in the form of
time--dependent Schr\"odinger equations, to be presented in Sect.
\ref{Gener}.

We first analyze the zero--temperature case (in Sect. \ref{Vacuum}),
where the relevant geometry is purely $AdS_5$. By solving the
`Schr\"odinger equations' in the approximations of interest and comparing
the results to the space--time picture of the quantum fluctuation of the
current into a pair of massless fields, we find that the two pictures
match indeed with each other after provided one identifies the radial
dimension $r$ in $AdS_5$ with the inverse $1/L$ of the transverse size of
the partonic fluctuation: more precisely, $R^2/r\leftrightarrow L$, wit
$R$ the curvature radius for $AdS_5$. For instance, on the supergravity
side, the temporal scale $t_c\sim q/Q^2$ appears as the time after which
the wave has penetrated into the bulk up to a distance $r\sim R^2Q$. This
corresponds, on the gauge theory side, to the fact that the partonic
fluctuation requires a formation time $t_c\sim q/Q^2$ in order to grow up
to a transverse size $L\sim 1/Q$. Notice that this formation time is a
genuine quantum effect, which reflects the uncertainty principle. This
suggests that, for the problem at hand, the 5th dimension of $AdS_5$
somehow mimics, within the context of the classical supergravity
calculation, the phase--space for physical quantum fluctuations in
transverse space.

This $r\leftrightarrow L$ duality helps us identify a physical picture
for the current dynamics in the vacuum. The naive one--loop picture
(2--parton fluctuation) cannot be right at strong coupling, where any of
the two partons produced in the original splitting of the current can
further radiate, or scatter off the vacuum fluctuations. Quantum field
are known for their strong propension to radiation. When the coupling is
weak, the radiation is concentrated in specific corners of the phase
space (collinear and soft radiation for a gauge theory), where the
smallness of the coupling is compensated by the phase--space available to
radiation. This leads, e.g, to the celebrated DGLAP and BFKL evolution
equations in perturbative QCD. But at strong coupling, there is no reason
why radiation should be restricted to small corners of phase--space; this
should most naturally proceed via the `democratical' branching of the
original quanta into two daughter partons with more or less equal shares
of their parent energy and momentum. We therefore propose a physical
picture for current (or parton) evolution a strong coupling (and in the
vacuum) in terms of successive branchings, such that the energy and the
virtuality are divided by two (for simplicity) at each individual
branching, and that the lifetime of a generation is determined by the
formation time for the next generation, in agreement with the uncertainty
principle. As we show in Sect. \ref{Phys}, this simple picture reproduces
indeed the results of the respective AdS/CFT calculation, at a
qualitative level.

Moving now to the finite--temperature case, we find (by solving the
appropriate `Schr\"odinger equations', in Sect. \ref{Plasma}) that, for
low and moderate energies, such that $q\ll Q^3/T^2$, and for not too
large values of time, the dynamics remains essentially as in the vacuum.
For a space--like current, this confirms the previous results in Ref.
\cite{Hatta:2007cs}: a `small meson', with size $L\sim 1/Q$, survives
almost unaltered in the plasma. (The `meson' can decay via tunnel effect,
but the decay rate is exponentially suppressed \cite{Hatta:2007cs}.) But
for a time--like current, the vacuum--like evolution proceeds only up to
a time $t_f \sim \sqrt{\gamma}/T$, with $\gamma=q/Q\gg 1$ the Lorentz
factor of the current. At this time $t_f$, the `current' --- or, more
precisely, the partonic system produced via its successive branching ---
has propagated inside the plasma over a longitudinal distance $z_f \sim
\sqrt{\gamma}/T$ and has extended in transverse space up to a size
$L_f\sim 1/\sqrt{\gamma}T$. Note that this value $L_f$ is precisely of
the order of the meson screening length computed in Refs.
\cite{Peeters:2006iu,Liu:2006nn,Chernicoff:2006hi}, and more commonly
written as $L_f\sim (1-v^2)^{1/4}/T$, with $v$ the velocity of the meson.
More detailed comparisons between our approach and previous results in
the literature will be performed in Sect. \ref{Paral}.

For larger times $t\gtrsim t_f$, this partonic system starts interacting
with the plasma, although its transverse size is still much smaller than
$1/T$. In this regime, the dual string calculation shows that the Maxwell
wave feels the attraction of the black hole and thus undergoes an
accelerated fall towards the horizon, that it gets close to after a time
of order $t_f$. When this radial dynamics is translated to transverse
space, via the $r\leftrightarrow L$ duality mentioned before, it implies
that the partons feel a constant, decelerating, force in the transverse
directions,
 \beq
 \frac{\dif p_\perp}{\dif t}\,\sim \,-T^2\,,
 \eeq
whose precise physical meaning remains a little mysterious to us (see the
discussion in Sect. \ref{Phys}). By including the action of this force on
the dynamics of branching, we will be able to show that {\em
medium--induced branching} is a likely physical scenario, which is
qualitatively consistent with the results of the AdS/CFT calculation. By
`medium--induced branching' we mean that the change in virtuality over
the lifetime of a partonic generation is due to the transfer of
transverse momentum to the plasma, at a constant rate $\sim T^2$ (see
Sect. \ref{Phys} for more details).

The same physical scenario --- an accelerated fall of the Maxwell wave
into the black hole, which physically corresponds to medium--induced
branching --- holds also in the high--energy regime at $q\gtrsim
Q^3/T^2$, for {\em both} space--like and time--like currents, but in this
regime, this scenario starts to apply much earlier, after a time $t_s\sim
q/Q_s^2(q)$ which is much shorter than the formation time $t_c\sim q/Q^2$
for an on--shell partonic fluctuation. (We recall that $Q_s\sim
(qT^2)^{1/3}$ is the plasma saturation momentum.) Thus, such an energetic
current disappears in the plasma before having the time to create `jets'
(on--shell partons). Within this scenario, we find a natural physical
explanation for the `drag force' experienced by partons in the plasma
(originally computed for a heavy quark probe in Refs.
\cite{Herzog:2006gh,Gubser:2006bz}), and also for the `trailing string
solution' (the string attached to an energetic heavy quark propagating
through the plasma $AdS_5$--Schwarzschild geometry
\cite{Herzog:2006gh,Gubser:2006bz}), that we recognize as the dual of the
enveloping curve of the spatial distribution of partons produced through
medium--induced branching. It remains as an interesting open problem to
understand whether this scenario can naturally accommodate also other
quantities computed within AdS/CFT, so like the jet quenching
\cite{Liu:2006ug}, or the transverse momentum broadening
\cite{CasalderreySolana:2007qw,CasalderreySolana:2006rq}.

One can succinctly summarize the previous discussion as follows: While a
highly virtual (and hence very small) {\em space--like} `meson', with
virtuality $Q\gg Q_s(q)$, can survive in the strongly--coupled plasma
(essentially without feeling the latter), this is not also the case for
the jets produced by the decay of a highly virtual {\em time--like}
current, which can only propagate over a longitudinal distance $z_f \sim
\sqrt{\gamma}/T$ before disappearing in the plasma. Less virtual (or more
energetic) partonic systems, for which $Q\ll Q_s(q)$, cannot form in the
first place : the virtual partons melt in the plasma over a time $t_s\sim
q/Q_s^2(q)$, which is too short to get on--shell. The dissipation
mechanism is {\em universal}, i.e., the same for all kind of partons ---
off--shell or on--shell, massive or massless, quarks, scalars, or gluons
--- and it consists in medium--induced branching.

Let us conclude this Introduction with a remark concerning the spatial
distribution of the hadrons produced via the decay of a time--like
current in the vacuum. For the concept of hadron to make sense, we
supplement the theory with an infrared cutoff $\Lambda$ and assume that
the current is `hard' relative to this cutoff : $|Q|\gg \Lambda$. Then
the decay of the current via successive branchings will produce a system
of $\sim |Q|/\Lambda$ hadrons with small transverse momenta
$\sim\Lambda$, which are isotropically distributed in the transverse
plane. (In the rest frame of the current, this distribution would be
spherically symmetric.) This observation is the `time--like' counterpart
of a previous result that, at strong coupling, there are no large--$x$
partons (i.e., no partons carrying a sizeable fraction of the hadron
longitudinal momentum) in the wavefunction of an energetic hadron
\cite{Polchinski:2002jw,Hatta:2007he,Hatta:2007cs}. Rather, all partons
have fallen down at very small values of $x$ (smaller than $x_s\sim T/Q$
in the case of the ${\mathcal N}=4$ SYM plasma \cite{Hatta:2007cs} and,
respectively, $x_s\sim (\Lambda/Q)^2$ for a hadron \cite{Hatta:2007he}),
via successive branchings within space--like cascades. Because of that, a
high--energy `nucleus--nucleus' collision which would liberate these
partons would produce no jets, but only a multitude of particles at
central rapidity which are isotropically distributed in transverse space,
with small momenta $p_\perp\sim\Lambda$. This picture is similar, and
possibly related, to a very recent result \cite{Hofman:2008ar} showing
that the energy distribution of the particle produced in the
strong--coupling analog of the $e^+e^-$ annihilation exhibits spherical
symmetry. On the other hand, this picture looks quite different from the
corresponding one in perturbative QCD and also from the respective
results for nucleus--nucleus collisions at RHIC.

%(in Sect. \ref{Vacuum})

\section{General equations}
\setcounter{equation}{0} \label{Gener}

We would like to address the following physical problem: at time $t=0$,
an ${\mathcal R}$--current with momentum $q$ oriented along the $z$ axis
and energy $\omega$ acts in the ${\mathcal N}=4$ SYM plasma, producing a
system of SYM quanta which then propagate through the plasma until they
finally disappear. In the dual gravity problem, this current dynamics is
represented by the propagation of a Maxwell--like gauge field $A_m$ in
the background geometry of the $AdS_5$ black hole (representing the
${\mathcal N}=4$ SYM plasma) \cite{Son:2007vk}. The corresponding metric
reads
 \beq \label{met1} \rmd s^2=\frac{(\pi TR)^2}{u}(-f(u)\rmd t^2+\rmd
 \bm{x}^2)+\frac{R^2}{4u^2f(u)}\rmd u^2\,,
 \eeq
where  $T$ is the temperature of the black hole (the same as for the
${\mathcal N}=4$ SYM plasma), $R$ is the curvature radius of $AdS_5$, $t$
and $\bm{x}=(x,y,z)$ are the time and, respectively, spatial coordinates
of the physical Minkowski world, $u$ is the radial coordinate on $AdS_5$,
and $f(u)=1-u^2$. Note that our radial coordinate has been rescaled in
such a way to be dimensionless: in terms of the more standard,
dimensionfull, coordinate $r$, it reads $u\equiv (r_0/r)^2$, with
$r_0=\pi R^2 T$. Hence, in our conventions, the black hole horizon lies
at $u=1$ and the Minkowski boundary at $u=0$.

Since the radial component $A_u$ vanishes at $u=0$, it is possible and
convenient to work in the gauge where $A_u$ is identically zero. Then,
the non--zero components $A_\mu$, with $\mu=0,1,2,3$, are of the form:
   \beq \label{pw}
   A_\mu(t,\bm{x},u)\,=\,\rme^{-i\omega t+iq z}\,\tilde A_\mu(t,u)\,, \eeq
where $\tilde A_\mu(t,u)$ has a relatively weak time dependence, such
that $|\del_t\tilde A_\mu|/\tilde A_\mu\ll \omega$, and obeys the initial
condition that, for $t\to 0$, $\tilde A_\mu(t,u)$ is localized near
$u=0$. (The precise structure of the initial condition is not really
needed.) The relevant equations of motion are the Maxwell equations in
the $AdS_5$ Schwarzschild geometry. They are most conveniently written as
a set of equations for the transverse components $\tilde A_i$, with
$i=1,2$, and, respectively, the longitudinal component $a(u)\equiv \del_u
A_0(u)$ (the $z$ component $A_z$ is not independent, but it is related to
$A_0$ via the equations of motion), and read \cite{Son:2007vk}
   \beq
   A_i^{\prime\prime}+\frac{f^\prime}{f} A_i^\prime\,
    - \,\frac{1}{uf^2}\,
   \left(k^2f +\frac{\del^2}{\del\tilde t^2}\right) A_i\,=\,0
   \label{ai} \eeq
and, respectively,
 \beq
a^{\prime\prime}+\,\frac{(uf)^\prime}{uf}\,a^\prime - \,\frac{1}{uf^2}\,
\left(k^2f +\frac{\del^2}{\del\tilde t^2}\right) a\,=\,0\,, \label{key}
   \eeq
where a prime on a field indicates a $u$--derivative and we have
introduced the following, dimensionless, variables
 \beq\label{dimko} \varpi\equiv
 \frac{\omega}{2\pi T}\,, \qquad  k\equiv \frac{q}{2\pi T}\,,
 \qquad \tilde t\equiv 2\pi T t\,,\qquad \tilde z\equiv 2\pi T z\,.
 \eeq
(The variables $\varpi$ and $\tilde z$ will be used later on.) The above
equations can be further simplified for our present purposes. First, as
already mentioned, the field $\tilde A_\mu(t,u)$ is slowly varying in
time; hence we can write
 \beq
 \frac{\del^2}{\del\tilde t^2}\,A_\mu\,\approx\,
 \rme^{-i\varpi \tilde t+ik \tilde z}\,\left(-\varpi^2-2i\varpi
 \frac{\del}{\del\tilde t}\right)\tilde A_\mu\,,\eeq
where we have neglected the second--order time derivative of $\tilde
A_\mu$. Furthermore, as we shall see, the interesting dynamics happens
near to the boundary at $u=0$ (and hence far away from the horizon at
$u=1$), so we can replace $f\to 1$ in the coefficients of the above
equations everywhere except in the term $k^2f=k^2(1-u^2)$ : indeed, in
that term, the small quantity $u^2\ll 1$ can be amplified by the
longitudinal momentum $k^2$ of the incoming current, which becomes very
large at high energy. Notice that this term $k^2 u^2=k^2(r_0/r)^4\propto
k^2T^4/r^4$ represents the potential for the long--range (in $r$)
gravitational interaction between the current and the black hole. Since
we keep only this particular medium effect in our equations, but neglect
those which would signal the black--hole singularity at $u=1$ (the
gravity--dual hallmark of a thermal system), we expect our subsequent
results to apply to matter distributions more general than a
finite--temperature plasma: similar results should hold for any matter
distribution which is infinite and homogeneous (say, a cold matter),
after replacing the energy density $\Theta_{00}\propto N_c^2T^4$ of the
finite--temperature plasma by the corresponding quantity for the cold
matter.

After these simplifications, we end up with the following equations for
$\tilde A_i$ and $\tilde a$, valid when $u\ll 1$ :
 \beq\label{tildeqs}
 2i\varpi \frac{\del}{\del\tilde t}\,\tilde A_i &\,=\,&
 -u \tilde A_i^{\prime\prime} \pm K^2 \tilde A_i -k^2u^2 \tilde A_i\,,\nn
 2i\varpi \frac{\del}{\del\tilde t}\,\tilde a &\,=\,&
 -u \tilde a^{\prime\prime} -\tilde a^{\prime} \pm K^2\tilde
 a -k^2u^2\tilde
 a\,,\eeq
where we have introduced the notation $K^2 \equiv |k^2-\varpi^2|$ and the
plus (minus) sign in front of $K^2$ corresponds to a space--like
(time--like) current. Note that, with its above definition, $K^2$ is
always a positive quantity.

It will be furthermore useful (especially in view of constructing
approximate solutions) to rewrite these equations in a form which
resembles the time--dependent Schr\"odinger equation. To that aim, we
shall perform the following changes of variable and functions:
 \beq\label{change}
 \chi\,\equiv\,2\sqrt{u},\qquad\tilde
 a(\tilde t,u)\,\equiv\,\frac{1}{\sqrt{\chi}}\,\psi(\tilde t,\chi),
 \qquad\tilde A_i \,\equiv\,{\sqrt{\chi}}\,\phi_i(\tilde t,\chi)\,.
 \eeq
Also, we shall be mostly interested in the high energy regime where
$\omega\simeq q\gg Q$, meaning $\varpi\simeq k\gg K$, so that we can
replace $\varpi\simeq k$ in the coefficients of the equations. Then our
final equations, valid for high energy and $\chi\ll 1$, read
 \beq
 i\frac{\del  \phi_i}{\del\tilde t} &\,=\,&\left(-\frac{1}{2k}\,
 \frac{\del^2}{\del \chi^2}\,+\,\frac{3}{8k\chi^2}\,\pm\,
 \frac{K^2}{2k}\,-\,\frac{k\chi^4}{32}\right)\phi_i\,,
 \label{SchT}\\
 i\frac{\del  \psi}{\del\tilde t} &\,=\,&\left(-\frac{1}{2k}\,
 \frac{\del^2}{\del \chi^2}\,-\,\frac{1}{8k\chi^2}\,\pm\,
 \frac{K^2}{2k}\,-\,\frac{k\chi^4}{32}\right)\psi\,. \label{SchL}
 \eeq
To avoid a proliferation of cases, we shall mostly focus on the
longitudinal case, cf. Eq.~(\ref{SchL}).

As anticipated, we choose initial conditions such that, at $t=0$, the
fields are localized near $u=0$ (i.e., $\chi=0$). We shall implement that
by working in Fourier space; we thus write, e.g.,
 \beq \psi(\tilde t,\chi)\,=\int \rmd\varepsilon \, \rme^{-i\varepsilon
 \tilde{t}}\,\psi(\varepsilon,\chi)\,,\eeq
where $\psi(\varepsilon,\chi)$ is a wave packet in $\varepsilon$ peaked
around $\varepsilon=0$. For convenience, we shall take this wave packet
to be modulated by a Gaussian, that is,
 \beq\label{gauss}
  \psi(\tilde t,\chi)\,=\int \rmd\varepsilon \, \rme^{-i\varepsilon
 \tilde{t}-\frac{\varepsilon^2}{2\sigma^2}}\,\Psi(\varepsilon,\chi)\,,\eeq
where the width $\sigma$ is constrained by
 \beq k \gg
 \sigma \gg \frac{K^2}{k}\, \label{gau} \eeq
and the additional $\varepsilon$--dependence in $\Psi(\varepsilon,\chi)$
will be fixed by Eq.~(\ref{SchL}). The first inequality in
Eq.~(\ref{gau}) ($k \gg \sigma$) ensures that the typical energy
fluctuations obey $\varepsilon\ll \varpi \simeq k$, as originally
assumed. The second inequality ($\sigma \gg {K^2}/{k}$) is necessary to
allow for quasi--localized configurations in the initial condition and at
early times (see Appendix A). This implies that the allowed fluctuations
can be quite large, $|\varpi-k|\sim {K^2}/{k}$, so that the space--like,
or time--like, nature of the current (depending upon the sign of
$\varpi-k$) becomes apparent only for sufficiently large times, after the
effects of the initial condition have dissipated. In fact, as we shall
see in the next section, the early--time behavior of the solution
$\psi(\tilde t,\chi)$ is independent of $K^2$, and hence the same for
both space--like and time--like currents.

Another boundary condition refers to the behavior of the solution in the
stationary regime (meaning, for large enough times) and at relatively
large values of $\chi$ : since a black hole is a purely absorptive
medium, from which no signal can escape, the solution $\psi(\tilde
t,\chi)$ near the horizon located at $\chi=2$ must be a purely outgoing
wave (i.e., a wave departing from the boundary and impinging in the black
hole). In fact, this outgoing--wave behavior should manifest itself
already for relatively low values $\chi\ll 1$, and hence can be used as a
boundary condition on Eq.~(\ref{SchL}), since the potential term in this
effective Schr\"odinger equation is monotonous everywhere except near the
boundary at $\chi=0$, and hence it cannot generate reflected waves. This
will be further discussed in the forthcoming sections.

\section{Jets in the vacuum}
\setcounter{equation}{0} \label{Vacuum}

We start with the zero--temperature case, i.e., with the propagation of
the Abelian current through the vacuum. The results to be obtained here
will not only serve as a level of comparison for the subsequent
discussion of a plasma, but they will also prepare that discussion, at
two levels (at least): First, as we shall see, there are special regimes
(so like early times) where the dynamics in the plasma is quite similar
to that in the vacuum. Second, the physical interpretation of our
results, whose understanding is the main purpose in this paper, is easier
to introduce and motivate in the context of the vacuum dynamics.

\subsection{The vacuum polarization tensor}
\label{Polar}

We begin this discussion with the {\em stationary case}, i.e., the case
where the fields are purely plane waves in the physical, 4--dimensional,
space, meaning that the fields denoted with a tilde in the previous
discussion (e.g., $\tilde A_\mu$) are independent of time. The
corresponding AdS/CFT calculation will provide the retarded
current--current correlator (or `polarization tensor') in Fourier space
(with $q^\mu=(\omega,0,0,q)$) :
   \beq
 R_{\mu\nu}(q)\,\equiv\,i\int \rmd^4x\,\rme^{-iq\cdot x}\,\theta(x_0)\,
 \langle [J_\mu(x), J_\nu(0)]\rangle\,, \label{Rdef}  \eeq
and in that sense it is the analog at strong coupling of computing
momentum--space Fourier diagrams in perturbation theory at weak coupling.
More precisely, $R_{\mu\nu}(q)$ is obtained by differentiating the
classical action with respect to boundary values of the fields at $u=0$ :
  \beq R_{\mu\nu}(q)\,=\,\frac{\partial^2 \mathcal{S}}{\partial A_\mu
 \partial A_\nu}\,,
 \label{SR} \eeq
where $A_\mu\equiv \tilde A_\mu(u=0)$ in the notations of Eq.~(\ref{pw})
and $\mathcal{S}$ is the four--dimensional action density, which is
homogeneous: $S=\int\rmd^4 x\, \mathcal{S} = \Delta V\,\Delta
t\,\mathcal{S}$, with $\Delta V\,\Delta t =$ the volume of space--time.
In turn, the classical action density can be fully expressed (after using
the equations of motion) in terms of the values of the field $\tilde
A_\mu(u)$ and of its first derivative at $u=0$ :
 \beq \mathcal{S} =\,\frac{N^2T^2}{16}\,
 \left[-\tilde A_0\partial_u \tilde A_0^* + f\tilde A_3 \partial_u \tilde A_3^*
  + f\tilde A_i \partial_u \tilde A_i^*
\right]_{u=0}
    \,, \label{actioncl} \eeq
where the $z$ component $\partial_u \tilde A_3$ is determined by the EOM
as $\partial_u \tilde A_3= -(\varpi/kf)\partial_u\tilde A_0$. A star on a
field denotes complex conjugation: the classical solutions develop an
imaginary part (in spite of obeying equations of motion with real
coefficients) because of the outgoing--wave condition at large $u$ (see
below). Via Eq.~(\ref{SR}), this introduces an imaginary part in
$R_{\mu\nu}(q)$ which physically describes the dissipation of the current
in the original gauge theory. In fact, the imaginary part of the
expression within the square brackets in Eq.~(\ref{actioncl}) is
independent of $u$ (as it can be checked by using the EOM) and hence it
can be evaluated at any $u$ \cite{Son:2002sd}.

Note that, even in this zero--temperature context, we keep using the
dimensionless variables $u$ and $\chi$, which were previously defined
with respect to the black--hole horizon $r_0=\pi R^2 T$. Throughout this
section, it will be understood that $r_0$ and $T$ are arbitrary (length
and, respectively, momentum) scales of reference, which are used to
construct dimensionless coordinates at intermediate stages of the
calculations, but which will drop out from the final, physical, results.
In this context, the radial coordinates $u$ and $\chi$ can take on all
the values from 0 to $\infty$.

We now turn to the actual calculation of the vacuum polarization tensor
for the ${\mathcal N}=4$ SYM theory in the strong coupling limit. The
outcome of this calculation is already known (the case of a space--like
current has been treated in detail in Ref. \cite{Hatta:2007cs}), so our
subsequent presentation will be very streamlined, with emphasis on the
physical interpretation of the results. We shall give details for the
longitudinal sector alone. The relevant equation reads (cf.
Eq.~(\ref{key}) in which we let $f\to 1$ and ${\del^2}/{\del\tilde t^2}
\to -\varpi^2$)
  \beq
 a^{\prime\prime}+\frac{1}{u}\,a^\prime\mp\frac{K^2}{u}\,a\,=\,0\,,
 \label{eqvac} \eeq
where we recall that the upper (lower) sign in front of $K^2 \equiv
|k^2-\varpi^2|$ corresponds to a space--like (time--like) current. After
the change of variable $ \chi\equiv 2\sqrt{u}$, this is recognized as the
equation for the $\nu=0$ Bessel functions, of either real, or imaginary,
argument (depending upon the sign in front of $K^2$). The solution is
constrained by the boundary condition (see Ref. \cite{Hatta:2007cs} for
details)
  \beq \lim_{u\to 0}\big[u a'(u)\big]\,=\,k(kA_0+\varpi
  A_3)\big|_{u=0}\,,
 %\,\equiv\,k^2\mathcal{A}_L(0)\,.
 \label{bc} \eeq
together with the condition of regularity at $u\to\infty$. For the
space--like case, these conditions uniquely determine the solution as
   \beq a(u)
  =-2k(kA_0+\varpi
  A_3)\big|_{u=0}\,\mathrm{K}_0(2K\sqrt{u})\qquad\mbox{(space--like)}
  \,. \label{a0SL} \eeq
The other independent solution, $\mathrm{I}_0(2K\sqrt{u})$, is rejected
since it would exponentially diverge as $u\to\infty$. For the time--like
case, on the other hand, the general solution is a superposition of
oscillating Bessel functions,
  \beq a(u)=c_1\mathrm{J}_0(2K\sqrt{u})+
  c_2\mathrm{N}_0(2K\sqrt{u})\,, \label{a0TLgen}\eeq
so the condition of regularity at $u\to\infty$ introduces no constraint.
To fix the solution in this case, we shall require $a(t,u)=\rme^{-i\omega
t}a(u)$ to be an {\em outgoing wave} at large $u$; then, the solution
becomes imaginary, with the appropriate sign for the imaginary part to
yield the {\em retarded} polarization tensor, via Eq.~(\ref{SR}). This
constraint implies $c_1=-ic_2$ which together with the boundary condition
(\ref{bc}) completely fixes the solution as
  \beq a(u)\,=\,-i\,{\pi}k(kA_0+\varpi
  A_3)\big|_{u=0}\,\mathrm{H}_0^{(1)}
   (2K\sqrt{u})\qquad\mbox{(time--like)}\,, \label{a0TL} \eeq
where $\mathrm{H}_0^{(1)}=\mathrm{J}_0+i\mathrm{N}_0$ is a Hankel
function encoding the desired outgoing--wave behavior at large $u$ :
$a(t,u)\propto \rme^{-i(\omega t-K\chi)}$ when $\chi\equiv 2\sqrt{u}\gg
1/K$.

The transverse--wave solutions can be similarly obtained, but the above
solutions for the longitudinal wave are in fact sufficient to complete
the calculation of $R_{\mu\nu}(q)$: by Lorentz and gauge symmetry, the
polarization tensor is transverse (with $\eta_{\mu\nu}=(-1,1,1,1)$) :
 \beq\label{R0} R_{\mu\nu}(q)=
  \left(\eta_{\mu\nu}-\frac{q_\mu q_\nu}{Q^2} \right)R(Q^2)\,\qquad
 \mbox{(vacuum)}\,,\eeq
and the scalar function $R(Q^2)$ can be computed with the longitudinal
waves alone. A standard calculation, which involves the removal of a
logarithmic divergence in the real part at $u=0$ (the AdS/CFT analog of
ultraviolet renormalization), finally yields
  \beq\label{Rvac}
 R(Q^2)\,=\,\frac{N^2_c|Q^2|}{32\pi^2}\left(\ln\frac{|Q^2|}{\mu^2}
 - i\pi\Theta(-Q^2)\rm{sgn}(\omega)\right)
 \,,\eeq
where $Q^2\equiv -\omega^2+q^2$ is the current virtuality in physical
units and $\mu$ is the renormalization scale. As expected, the imaginary
part is non--zero only for a time--like ($Q^2<0$) current, which can
decay into the massless fields (adjoint scalars and Weyl fermions) of the
${\mathcal N}=4$ SYM theory carrying ${\mathcal R}$--charge.

As anticipated in the Introduction, this result (\ref{Rvac}) is exactly
the same as the respective perturbative result to one--loop order, so in
particular its imaginary part (formally) describes the current decay into
one pair of massless fields (a quark--antiquark pair, or two scalar
fields). This interpretation is, of course, only formal: At strong
coupling, the current can couple to arbitrarily complicated
multi--particle states, but it so happens that, for the ${\mathcal N}=4$
SYM theory, the total cross--section is determined by the two--particle
final states alone. Our present calculation being an `inclusive' one ---
it provides the total cross--section, but it does not discriminate
between the various final states ---, its result cannot give us any
direct insight into the nature of these final states. We shall later try
to gain such an insight based on physical considerations.

We conclude this discussion of the stationary case with another point of
physical interpretation, which points towards an interesting `duality'
between the radial dimension in $AdS_5$ and the transverse size of the
current (or, more precisely, of the partonic system into which the
current has evolved) in the physical Minkowski space. Consider a
space--like current, for definiteness. The modified Bessel function in
Eq.~(\ref{a0SL}) decays exponentially when $2K\sqrt{u}\gg 1$, meaning
that the current penetrates in the radial dimension only up to a finite
distance $u_0\sim 1/4K^2$, or $\chi_0\sim 1/K$ (recall that $\chi\equiv
2\sqrt{u}$). The more virtual the current is, the closer it remains to
the boundary. This is quite similar to the picture of the current in
transverse space, as familiar in perturbation theory (for either QCD or
${\mathcal N}=4$ SYM): the virtual current fluctuates into a
quark--antiquark pair (a `color dipole') whose transverse size is
inversely proportional to the current virtuality: $L\sim 2/Q$. This
analogy is in fact even closer: the longitudinal wave solution in
Eq.~(\ref{a0SL}) involves the same modified Bessel function as the
wavefunction describing the dipole fluctuation of a longitudinal photon
in lowest order perturbation theory, which allows us to identify the
respective arguments as $K\chi\leftrightarrow QL/2$. Recalling that
$K=Q/2\pi T$ for a space--like current, we deduce the correspondence
$\chi\leftrightarrow \pi T L$. The same argument holds in the transverse
sector, where $\mathrm{K}_0$ is replaced by $\mathrm{K}_1$. Also, the
argument can be adapted to a time--like current, since the oscillatory
Bessel functions in Eq.~(\ref{a0TL}) are rapidly decaying at radial
distances $u\gg 1/4K^2$.

The above writing of the correspondence between the transverse size of
the partonic fluctuation and the radial coordinate in $AdS_5$, namely,
$\chi\leftrightarrow \pi T L$, explicitly involves the temperature and
hence it might look a bit formal in the present context of the vacuum
(but this writing will be natural for the subsequent discussion of a
plasma). To avoid confusion, it is preferable to recall the definition
$\chi\equiv 2(r_0/r)$ with $r_0\equiv \pi R^2 T$ in order to rewrite this
correspondence in the form $2R^2/r \leftrightarrow L$, from which the
temperature has dropped out.

To summarize the previous discussion, the vacuum polarization tensor in
the ${\mathcal N}=4$ SYM theory at strong coupling formally describes the
fluctuation of the current into a pair of elementary, massless, fields,
whose transverse size appears to be in a one--to--one correspondence with
the radial distance for the current penetration in $AdS_5$. As we shall
see, this interpretation is comforted by the discussion of the
time--dependent case, to which we now turn.

\subsection{Jet evolution in the vacuum}
\label{Coher}

As explained in Sect. \ref{Gener}, we are interested in the time
evolution of vector fields $\tilde A_\mu(t,u)$ which at $t=0$ start as a
perturbation localized near $u=0$ (or $\chi=0$). Given the correspondence
between $\chi$ and $L$ (the transverse size of the current), as argued at
the end of the previous subsection, we see that this initial condition
corresponds to a current which, at $t=0$, is point--like in transverse
space. To describe its evolution, we shall use the Schr\"odinger form of
the equations of motion, cf. Eqs.~(\ref{SchT})--(\ref{SchL}), and focus
on the longitudinal sector, for definiteness. The vacuum version of these
equations is obtained by removing the last term $\propto \chi^4$ in the
potential; this yields, e.g.,
 \beq
i\frac{\del  \psi}{\del\tilde t} &\,=\,&\left(-\frac{1}{2k}\,
 \frac{\del^2}{\del \chi^2}\,-\,\frac{1}{8k\chi^2}\,\pm\,
 \frac{K^2}{2k}\right)\psi\,,
  \label{Lvac}\eeq
for which we shall construct solutions obeying the relevant initial and
boundary conditions. To remain as simple as possible, we shall consider
approximate solutions which are valid piecewise in $\chi$ and which are
sufficient to illustrate the main points of physics. A more systematic
method to construct solutions to Eq.~(\ref{Lvac}), which is based on the
wave--packet decomposition in Eqs.~(\ref{gauss})--(\ref{gau}), will be
described in Appendix A.

We first note that at early times, so long as $\chi$ remains smaller than
a critical value $\chi_c\simeq 1/2K$ (the corresponding limit on time
will be determined later on), the second term, proportional to $K^2$, in
the potential in Eq.~(\ref{Lvac}) can be neglected compared to the first
term $\propto 1/\chi^2$. Thus, this early--time dynamics is identical for
both space--like and time--like currents. With the $K^2$ term omitted,
Eq.~(\ref{Lvac}) admits the following, exact, solution
 \beq\label{psivac}
 \psi(\tilde t, \chi)\,=\,-i
  \frac{\sqrt{\chi}}{\tilde{t}}\
  \rme^{i\frac{k\chi^2}{2\tilde{t}}}\,. \eeq
This implies that the actual longitudinal wave (cf. Eq.~(\ref{change}))
 \beq
 \tilde a(\tilde t,\chi)\,=\,\frac{1}{\sqrt{\chi}}\,\psi\,\propto\,
  \frac{1}{\tilde{t}}\
 \rme^{i\frac{k\chi^2}{2\tilde{t}}}\,, \label{we} \eeq
is localized near $\chi=0$ at $\tilde{t}=0$ (although this is not exactly
a delta function). With increasing time, the energy density carried by
the wave (\ref{we}) diffuses towards larger values of $\chi$, so that the
typical distance traveled by the corresponding wave--packet after a time
$\tilde{t}$ is
 \beq \chi_{\rm diff}(\tilde t)
 \,\simeq\, \sqrt{\frac{2\tilde{t}}{k}}
 %\,\qquad\mbox{when}
 %\qquad \tilde t\,\ll\,\frac{k}{K^2}
 \,. \label{dif}
 \eeq
This behavior holds so long as $\chi_{\rm diff}(\tilde t)\lesssim 1/2K$,
meaning for times $\tilde t\lesssim \tilde t_c\sim k/K^2$. In physical
units, this yields $t_c\sim q/|Q^2|$, which is precisely the {\em
coherence time} of the high--energy current; that is, this is the time
interval which controls the Fourier transform\footnote{At least in the
vacuum, i.e., in the absence of other time scales which are introduced by
a medium. As we shall see in Sect. \ref{Plasma}, a finite--temperature
plasma introduces a new such a scale indeed.} in Eq.~(\ref{Rdef}), as it
can be checked by rewriting the complex exponential there as
 \beq\label{exp}
 \rme^{-i\omega t+iqz}\,\simeq\,\rme^{-iq(t-z)+iQ^2t/2q}\,, \eeq
where we have used $\omega\simeq q - Q^2/2q$ at high energy.

It is interesting to compare these results with the evolution of the
quark--antiquark (`dipole') fluctuation of a virtual photon in
perturbation theory in QCD: if the photon dissociates at $t=0$ into a
point--like pair of fermions, then with increasing time the transverse
size of this pair is increasing diffusively, due to quantum dynamics,
like \cite{Farrar:1988me,Dokshitzer:1991wu}
 \beq L\sim \sqrt{\frac{t}{q}}\,, \eeq
until it reaches a maximal size $L\sim 1/\sqrt{|Q^2|}$ at a time $t_c\sim
q/|Q^2|$. (To avoid cumbersome notations, we shall often write $Q$
instead of $\sqrt{|Q^2|}$ and $Q^2$ instead of $|Q^2|$ within parametric
estimates. For instance, the coherence time will be estimated as $t_c\sim
q/Q^2$, where the modulus on $Q^2$ is implicit for a time--like current.)
For $t > t_c$, the pair is either recombining back into a photon, or ---
if the photon was time--like --- it splits apart, thus giving rise to two
on--shell particles which move away from each other.

Clearly, the early time ($t < t_c$) evolution of the $q\bar q$
fluctuation in perturbation theory is very similar to the corresponding
evolution of the vector perturbation in $AdS_5$ provided one identifies
$\chi\sim TL$, in agreement with the discussion at the end of Sect.
\ref{Polar}. As we show now, this correspondence persists also for larger
times $t > t_c$.

Indeed, for $\tilde t  \gg \tilde t_c$, one can heuristically estimate
the time--derivative in the l.h.s. of Eq.~(\ref{Lvac}) as ${\del
}/{\del\tilde t}\ll 1/\tilde t_c \sim K^2/k$ (this heuristic argument
will be confirmed by the wave--packet analysis in Appendix A). That is,
the time--derivative term in the equation is much smaller than the last
term $\sim K^2/k$ in the potential, which becomes the dominant term when
$\chi\gg 1/2K$. Hence, for $\tilde t \gg \tilde t_c$ and $\chi\gg
\chi_c$, the equation simplifies to
 \beq
 \frac{\del^2}{\del\chi^2}\,\psi\,=\,\pm\,
 {K^2}\psi\,,
  \label{LLvac}\eeq
(i.e., Schr\"odinger equation in a flat potential) with the obvious,
acceptable, solutions
 \beq\label{psivaclarge}
    \psi(\chi)\,
     \propto\,
    \begin{cases}
        \displaystyle{\rme^{-K\chi}}
         &
        \text{ (space--like)}
        \\*[0.2cm]
        \displaystyle{\rme^{iK\chi}} &
        \text{(time--like)}
        \,.
    \end{cases}
    \eeq
which are time--independent and coincide, as they should, with the
asymptotic versions (valid at large $\chi\gg 1/K$) of the respective
stationary solutions constructed in Sect. \ref{Polar}. In the space--like
case, the above solution confirms that, for times $t > t_c$, the
perturbation remains localized near the boundary, within a distance
$\chi\lesssim 1/K$. In the time--like case, it implies that the actual
wave
 \beq\label{largetvac}
 a(\tilde t,\tilde z,\chi)\,=\,\rme^{-i\varpi \tilde t+ik \tilde z}\,
 \frac{\psi}{\sqrt{\chi}}\,\propto\,
  \exp\left\{-ik(\tilde t-\tilde z)\,
 -i\frac{K^2}{2k}\,\tilde t +iK\chi\right\}
 \eeq
propagates with constant group velocity $v_g=K/k\ll 1$ along the radial
direction of $AdS_5$ :
 \beq \label{groupvac}
 \frac{\del }{\del K}\left(\frac{K^2}{2k}\,\tilde t
 -K\chi\right)\,=\,0\qquad\Longrightarrow\qquad\chi_g(\tilde t)\,=\,
 \frac{K}{k}\,\tilde t\,.
 \eeq
Note that, for $\tilde t \sim \tilde t_c$, we have $\chi_g \sim 1/K$, as
it should for consistency with the previous solution at early times.
Hence, after the wave packet has diffused up to a distance $\chi_c \sim
1/K$ inside the bulk of $AdS_5$, it furthermore propagates with constant
radial velocity, so like a free particle.

This free--motion pattern at $t>t_c$ looks, of course, natural, in view
of the flatness of the potential in Eq.~(\ref{LLvac}). In Appendix B, we
shall verify that Eq.~(\ref{groupvac}) is the same as the radial part of
the geodesics of a massless classical particle which propagates in
$AdS_5$ with longitudinal velocity $v_z=q/\omega$ and radial velocity
$v_\chi=Q/\omega$, which is the same as the group velocity in
Eq.~(\ref{groupvac}) (recall that $\omega\simeq q$); this classical
particle is massless since $v_z^2+v_\chi^2=1$. On the other hand, the
classical particle dynamics cannot reproduce the diffusion at early
stages (a genuinely quantum effects), nor the fall of the wave into the
black hole (to be later described, in Sect. \ref{Fall}).

Via the correspondence $\chi\leftrightarrow \pi T L$, the result in
Eq.~(\ref{groupvac}) is again consistent with the transverse dynamics
expected for a time--like current which dissociates into a pair of
massless particles (in lowest--order perturbation theory). Indeed, this
result translates into
 \beq\label{Lfree}
 L(t)\,\simeq\,2\,\frac{\sqrt{\omega^2-q^2}}{\omega}\,t\,=\,
 2\,\sqrt{1-v_z^2}\,t\,=\,
 2\,v_\perp t\,,
 \eeq
where $v_z=q/\omega$ is the common longitudinal velocity of the two
massless particles, as inherited from the current, and $v_\perp=
\sqrt{1-v_z^2}$ is the modulus of their transverse velocity: the two
particles move in opposite directions in the transverse plane, so the
transverse distance between them increases like $L=2v_\perp t$. The above
results have been obtained by working in the high--energy regime where
$\omega\simeq q\gg Q$. However, it is easy to repeat the analysis for
other regimes, with similar conclusions. For instance, for a
zero--momentum current ($q=0$, $Q=\omega$), one finds that $\chi_g(\tilde
t)=\tilde t$, i.e., the two particles move in the transverse plane at the
speed of light ($v_\perp=1$).

At this point, one should again stress that the reason why the physical
picture looks so simple in the strongly--coupled ${\mathcal N}=4$ SYM
theory is because of the non--renormalization property of the
current--current correlator, as alluded to before. The AdS/CFT
calculation correctly provides the total cross--section for the current
decay, but it does not capture the detailed nature of the final states.
Formally, the total cross--section is saturated by the two--particle
final state; therefore, it is the dynamics of this particularly simple
state which emerges, via the correspondence $\chi\leftrightarrow \pi T
L$, from the dual calculation on the gravity side.

This discussion has an interesting corollary: it shows that the current
can be also viewed as a device for introducing a pair of elementary,
massless, fields of the ${\mathcal N}=4$ SYM theory (Weyl fermions or
adjoint scalars) at a given radial distance within $AdS_5$, which is
controlled by the current virtuality: $\chi \sim 1/K$ or $r\sim QR^2$.
This is tantamount to fixing the transverse size $L\sim 2/Q$ of the
partonic pair in the physical space. That is, the dynamics of the current
at times larger than the coherence time is the same as that of `meson',
or of a `color dipole'. For a space--like current, this effective `meson'
simply sits at $\chi \sim 1/K$, meaning that its transverse size is
fixed. For a time--like current, this `meson' propagates with constant
velocity along the radial direction $\chi$, meaning that its transverse
size grows at constant speed. In the next section we shall study the
influence of a thermal bath on this dynamics.

\section{Jets in the plasma}
\setcounter{equation}{0} \label{Plasma}

We now turn to the problem of main interest for us here, which is the
propagation of the current and of the associated partonic system through
a strongly--coupled ${\mathcal N}=4$ SYM plasma with temperature $T$. We
are interested in `hard probes', so we shall choose a current with
relatively high virtuality (either space--like, or time--like) : $Q\equiv
\sqrt{|Q^2|}\gg T$ (or $K\gg 1$), which probes the structure of the
plasma on distances much shorter than the thermal wavelength $1/T$. We
shall mostly consider a relativistic current, for which $q\simeq\omega\gg
Q$, but the non--relativistic case ($q\ll \omega$) will be briefly
discussed too, for completeness. In fact, the physically most interesting
case --- the one where the medium effects should be truly relevant --- is
when the coherence time $t_c\sim q/Q^2$ of the current is much larger
than $1/T$, so that the current explores a relatively large longitudinal
slice of the plasma, with width $\Delta z\sim t_c\gg 1/T$. This implies a
lower limit on the current momentum: $q\gg Q^2/T$, which is tantamount to
the condition that the Bjorken--$x$ variable\footnote{This variable is
especially relevant for a space--like current which undergoes deep
inelastic scattering off the plasma \cite{Hatta:2007cs}.}, defined as
$x\equiv Q^2/2qT$ (in the plasma rest frame), be very small: $x\ll 1$.

\subsection{Physical regimes}

 We would like to determine the characteristic time scale for
the dissipation of the current in the plasma; in the dual, gravity,
problem, this is the time scale for the fall--off of the Maxwell field
$A_\mu$ into the black hole. As we shall see, this scale is controlled by
the dynamics at relatively small $u\ll 1$, and thus is insensitive to the
detailed geometry of the black hole near its horizon at $u=1$. As
explained in Sect. \ref{Gener}, the strength of the gravitational
interactions between the wave and the black hole is proportional to the
wave longitudinal momentum $q$, and also to the temperature. In view of
that, we shall be led to distinguish between two important physical
regimes:

\texttt{(i)} a relatively low--energy (or low--temperature) regime at
$qT^2\ll Q^3$ (or $k\ll K^3$), where the medium effects are strongly
delayed, so that the current dynamics proceeds as in the vacuum up to
time scales much larger than $t_c$, and

\texttt{(ii)} a very high--energy (or high--temperature) regime at
$qT^2\gg Q^3$ (or $k\gg K^3$), in which the current dissipates very fast,
on a time scale much shorter than $t_c$.

One can understand these various regimes by studying the potential in the
Schr\"odinger--like equations (\ref{SchT})--(\ref{SchL}). Let us first
recall from the previous section that the time--dependence in these
equations is important only at very early stages, when the typical values
of $\chi$ are so small that the only relevant term in the potential is
the first term, $\propto 1/k\chi^2$, which is independent of both the
current virtuality and the properties of the medium. Hence, for such
early times, the equations describe diffusion, for both space--like and
time--like currents, and in the same way as in the vacuum. But for later
times, where the meaning of `later' is generally medium-- and
energy--dependent (see below), the solutions approach a stationary
regime, where Eqs.~(\ref{SchT})--(\ref{SchL}) can be replaced by their
time--independent versions, of the generic form $-\psi''+V\psi=E\psi$
with $E=0$ (since $E$ corresponds to the time--derivative, which is
negligible). Hence, this late--time behavior is determined by the
time--independent Schr\"odinger solution with zero energy in the
potential $V(\chi)$.

\FIGURE[t]{
\includegraphics[height=8.cm]{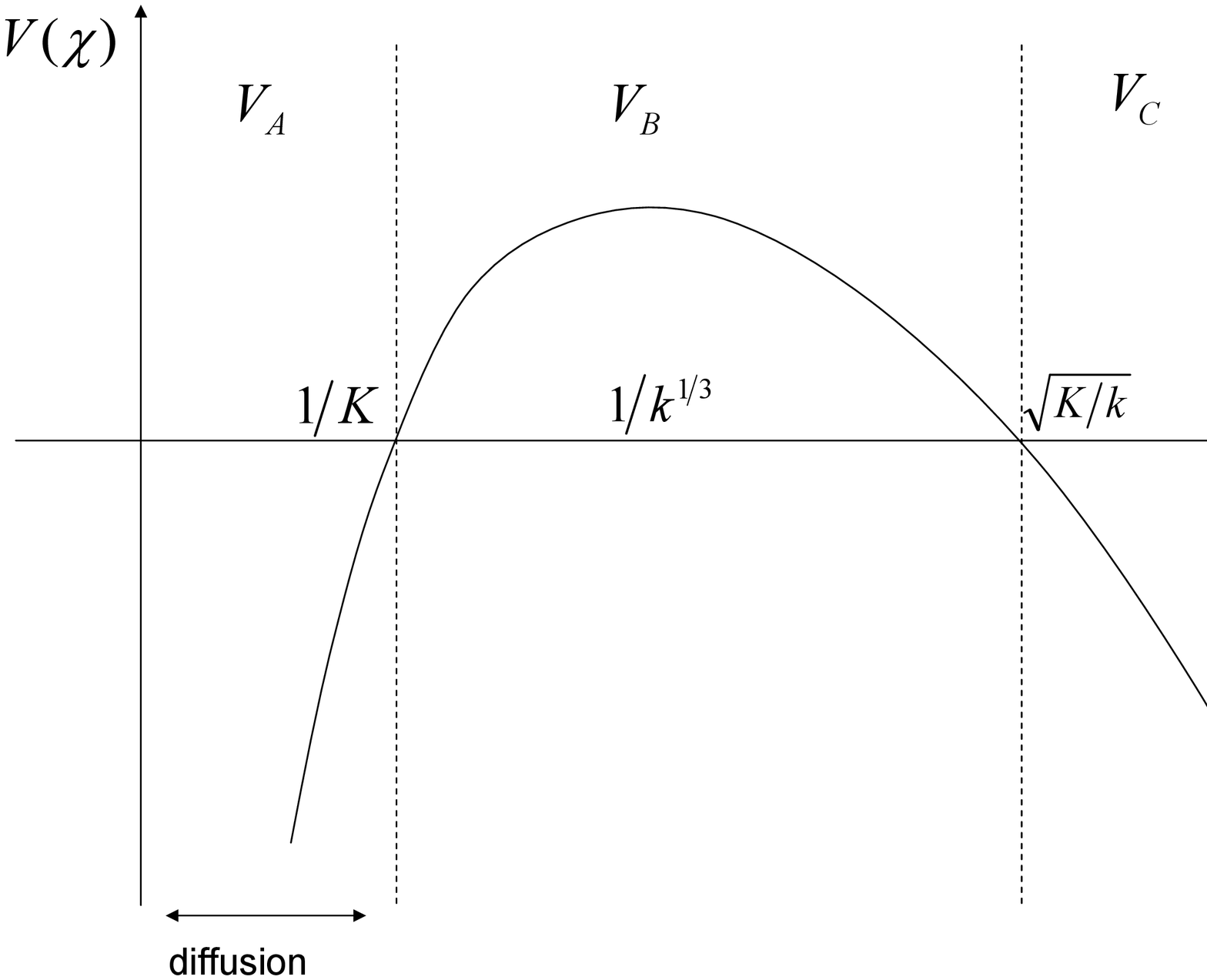}
\caption{\sl The potential in Eq.~(\ref{SchL}) for the space--like case
at relatively low energy ($k\ll K^3$). \label{PSlow}} }

\FIGURE{
\includegraphics[height=8.cm]{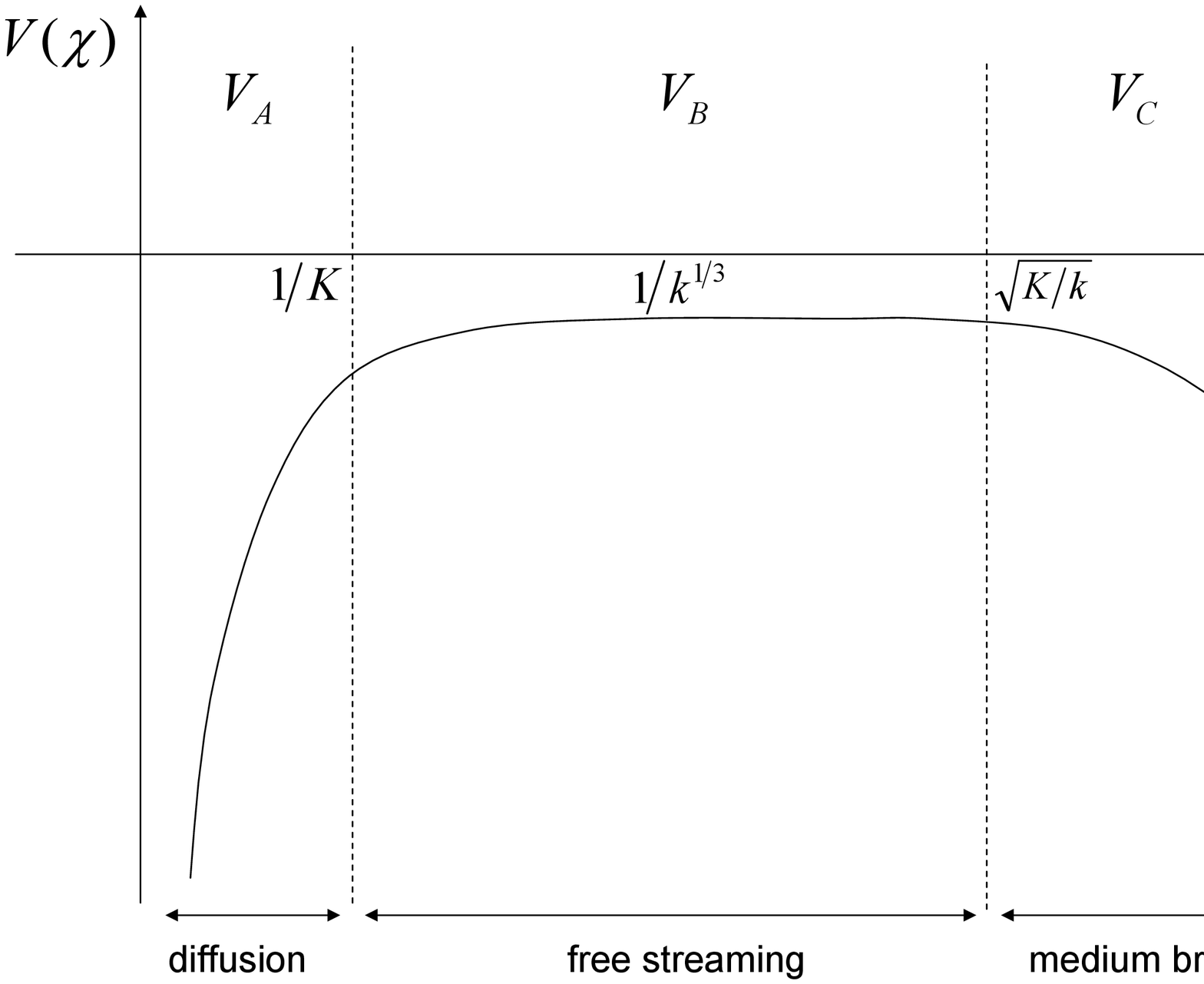}
\caption{\sl The potential in Eq.~(\ref{SchL}) for the time--like case at
relatively low energy ($k\ll K^3$). \label{PTlow}} }

\FIGURE[t]{
\includegraphics[height=8.cm]{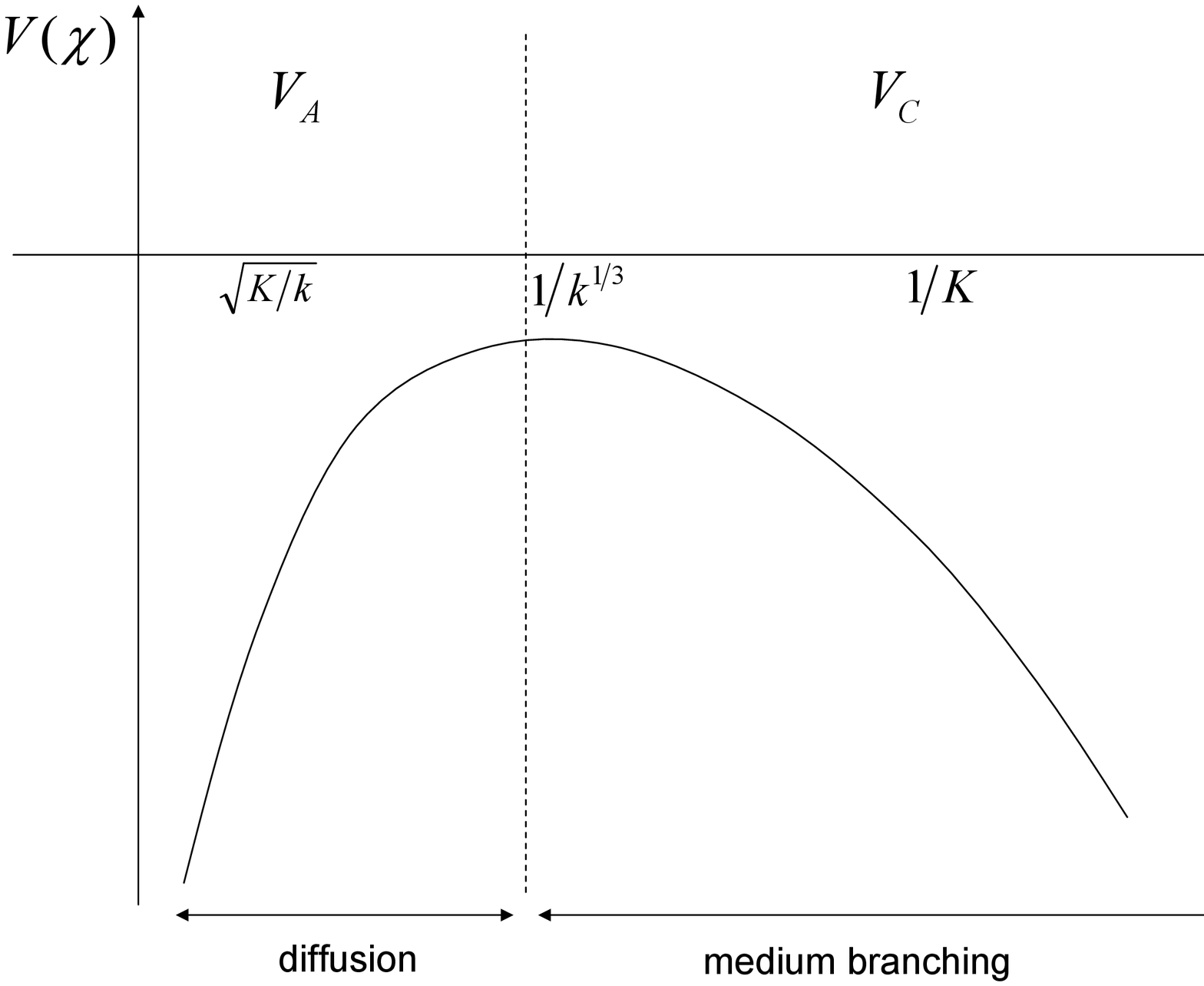}
\caption{\sl The potential in Eq.~(\ref{SchL}) in the high--energy regime
at $k\gg K^3$, where this potential is essentially the same for both
space--like and time--like currents. \label{Phigh}} }

In what follows we shall study this solution in the longitudinal sector
alone (the corresponding discussion for the transverse sector being very
similar). The respective potential is graphically illustrated in Figs.
\ref{PSlow}, \ref{PTlow}, and \ref{Phigh}, for the various physical
regimes: low energy and space--like in Fig. \ref{PSlow}, low energy and
time--like in Fig. \ref{PTlow}, and high energy (both space--like and
time--like) in Fig. \ref{Phigh}. Some general features of the dynamics
are already clear by inspection of these figures: At sufficiently small
values of $\chi$, the potential is the same as in the vacuum; the medium
starts to be felt only at relatively large values of $\chi$, where the
potential is a decreasing function which describes attraction by the
black hole. Also, in the high--energy regime, the dynamics is essentially
the same for both space--like and time--like currents. Finally, in the
space--like case at least, there is an obvious difference between the
low--energy regime, where the potential barrier constrains the wave to
remain on the left of the `classical turning point' at $\chi\sim 1/K$ (so
like in the vacuum), and the high--energy regime, where the barrier has
disappeared and the wave can easily propagate towards the horizon. (The
transition between the two regimes occurring at $k\sim K^3$ corresponds
precisely to the situation where the height of the potential barrier
becomes negligible.) In the time--like case, on the other hand, there is
never such a barrier, so the difference between the low--energy and
high--energy regimes looks perhaps less obvious. As we now explain, this
difference is nevertheless important in that case too.

To that aim, it is convenient to separate the three pieces in the
potential in Eq.~(\ref{SchL}), which play different physical roles. We
thus write (for a time--like current)
 \beq\label{VABC}
 V(\chi)\,=\,
 -\,\frac{1}{8k\chi^2}\,-\,
 \frac{K^2}{2k}\,-\,\frac{k\chi^4}{32}\,\equiv\,V_A+V_B+V_C\,.\eeq
Let $\chi_{AB}$ denote the value of $\chi$ at which $V_A$ and $V_B$
become comparable with each other, and similarly for $\chi_{AC}$ and
$\chi_{BC}$. We have, parametrically,
 \beq\label{chis}
 \chi_{AB}\,\sim\,\frac{1}{K}\,,\qquad
 \chi_{AC}\,\sim\,\frac{1}{k^{1/3}}\,,\qquad
 \chi_{BC}\,\sim\,\sqrt{\frac{K}{k}}\,.\eeq
Consider now the two physical regimes alluded to before:

\texttt{(i)} {\sf Low energy regime :} \ $k\ll K^3$ \ $\Longrightarrow$ \
$\chi_{AB}\,\ll\,\chi_{AC}\,\ll\,\chi_{BC}$ \\
It is easy to check that, in this regime, the potential admits the
piecewise approximation
 \beq\label{Vlow}
    V(\chi)\,
     \propto\,
    \begin{cases}
        \displaystyle{V_A+V_B}
         &
        \text{for $\chi\ll \chi_{BC}$}
        \\*[0.2cm]
        \displaystyle{V_C} &
        \text{for $\chi\gg \chi_{BC}$}
        \,.
    \end{cases}
    \eeq
Thus, so long as the relevant values of $\chi$ (those where most of the
wave energy is located) remain much smaller than $\chi_{BC}$, the
dynamics is the same as in the vacuum. This is the situation at
sufficiently small times, that we shall study in the next subsection. But
when $\chi\gtrsim \chi_{BC}$, the medium effects become important and
entail the fall of the wave into the black hole. This fall be studied in
Sect. \ref{Fall}. Note that the transition point $\chi\sim \chi_{BC}$ is
where the potential changes from a plateau to a rapidly decreasing
function (see Fig. \ref{PTlow}).

\texttt{(ii)} {\sf High energy regime :} \ $k\gg K^3$ \ $\Longrightarrow$
\ $\chi_{BC}\,\ll\,\chi_{AC}\,\ll\,\chi_{AB}$ \\
In this regime, the virtuality--dependent piece $V_B$ is never important,
and the potential can be approximated as (see also Fig. \ref{Phigh})
 \beq\label{Vhigh}
    V(\chi)\,
     \propto\,
    \begin{cases}
        \displaystyle{V_A}
         &
        \text{for $\chi\ll \chi_{AC}$}
        \\*[0.2cm]
        \displaystyle{V_C} &
        \text{for $\chi\gg \chi_{AC}$}
        \,,
    \end{cases}
    \eeq
which now holds for both space--like and time--like currents. Unlike the
low--energy potential, the high--energy one has no intermediate plateau,
but only a pronounced peak at $\chi\sim \chi_{AC}$. In fact, the
transition between the two energy regimes, which occurs for $k\sim K^3$,
corresponds to the situation where the two endpoints of the plateau,
$\chi_{AB}$ and $\chi_{BC}$, merge with each other, and also with the
position $\chi_{AC}$ of the emerging peak.

\subsection{Early--time dynamics}

In this subsection, we shall consider the dynamics of the current at
early times, which is insensitive to medium effects (except for its
validity limits, as introduced by the plasma), and thus can be inferred
from the corresponding discussion in Sect. \ref{Vacuum}.

We start with a simple case, which has not been explicitly covered by the
previous discussion, but which can be viewed as a special limit of the
low--energy case: a non--relativistic, time--like, current with $q\ll
\omega$ and $\omega\gg T$. The corresponding `Schr\"odinger' equations
are obtained from Eqs.~(\ref{SchT})--(\ref{SchL}) by omitting the last,
medium--dependent, term in the potential (since this term is
comparatively small for any $0\le\chi\le 2$) and replacing $k\to\varpi$
and $K^2\to\varpi^2$ in all the other terms. The ensuing equations are
the same as in the vacuum, and therefore so is also the current dynamics,
until it eventually dissipates into the plasma. Specifically, at very
early times $\tilde t\lesssim 1/\varpi$, the associated Maxwell wave
diffusively penetrates into $AdS_5$, up to a distance $\chi\sim 1/\varpi$
(meaning that the current diffusively spreads out in the physical space,
up to a size $L\sim 1/\omega$). Then, for larger times $\tilde t \gg
1/\varpi$, the wave propagates inside $AdS_5$ with constant group
velocity $v_g=1$, meaning that the current has decayed into a pair of
massless particles which move away from each other at the speed of light:
$\chi_g(\tilde t)=\tilde t$, or $L(t)=2t$. This behavior continues until
the wave hits the horizon ($\chi_g=2$ or $\tilde t\sim 1$) and is thus
absorbed by the black hole. Physically, this means that, for times $t\sim
1/T$, the separation between the products of the current decay has become
as large as the thermal wavelength, $L \sim 1/T$, so these products
cannot be anymore distinguished from the plasma fluctuations --- they get
`lost' in the plasma.

%The current disappears not because it was physically absorbed by the
%plasma (say, as a result of some scattering process), but because it has
%decayed
%--- in the same way as it would do in the vacuum --- and then the
%products of this decay were `lost' within the plasma.

We now turn to a relativistic current with $\varpi\sim k\gg K^2$, for
which we have to distinguish between two physical regimes, as previously
explained:

 \texttt{(i)} At moderate energies, such that $K^2\ll k\ll
K^3$, and for a time--like current, the upper limit $\chi_{BC}\sim
\sqrt{{K}/{k}}$ for the vacuum--like behavior is much larger than
$\chi_{AB}\sim{1}/{K}$, which is the maximal penetration length through
diffusion. Hence, in this regime, one can observe both stages of the
evolution identified in Sect. \ref{Vacuum}\,: a `diffusive' stage at
early times $\tilde t\lesssim \tilde t_c\sim k/K^2$, during which the
wave--packet progresses according to Eq.~(\ref{dif}), and a `free motion'
stage at $\tilde t > \tilde t_c$, during which the wave has constant
group velocity $v_g=K/k$. This second stage lasts until the position
$\chi_g(\tilde t)=v_g \tilde t$ of the wave--packet becomes comparable to
$\chi_{BC}$\,; this happens at a time $\tilde t_f\sim \sqrt{{k}/{K}}$
(the subscript $f$ on $\tilde t_f$ stands for ``free motion''), which
represents the upper time limit for this vacuum--like dynamics. In
physical units, this yields a time scale
 \beq\label{tg}
 t_f\,\sim\,\frac{1}{T}\,
 \sqrt{\frac{q}{Q}}\ \gg\ t_c\,\sim\, \frac{q}{Q^2} \ \gg\ \frac{1}{T}\,,
 \eeq
which, as indicated above, is much larger than the coherence time.

The physical interpretation of this result, as deduced via the
correspondence $\chi\leftrightarrow \pi T L$, is as follows: the
time--like current develops a partonic fluctuation with size $L\sim 1/Q$
over a time of the order of the coherence time $t_c$, and then decays
into a system of massless particles which move freely (without feeling
the plasma) up to a time $t\sim t_f$. At this stage, the transverse
extent of the partonic system has increased up to a value
 \beq\label{Lg}
 L_f\,=\,2v_\perp t_f
 \,\sim\,\frac{1}{T}\,
 \sqrt{\frac{Q}{q}} \ \ll\ \frac{1}{T}\,,
 \eeq
which is still small as compared to the thermal wavelength in the medium.
For even later times $t > t_f$, the dynamics is driven by the medium and
will be analyzed in the next subsection.

It is interesting to notice at this point that, although it influences
the dynamics at late times $t > t_f$, the plasma has no effect on the
polarization tensor, which is still given by the same expression as in
the vacuum, namely Eqs.~(\ref{R0})--(\ref{Rvac}) with $Q^2<0$. Indeed, if
one considers a (time--like) plane--wave solution, so like in Sect.
\ref{Polar}, then for $\chi\ll\chi_{BC}$ this solution is again given by
Eq.~(\ref{a0TL}), which describes an outgoing wave for any $\chi$ within
the range  $\chi_{AB}\ll\chi\ll \chi_{BC}$. The change in the solution at
larger $\chi\gtrsim \chi_{BC}$ has no incidence on the calculation of
$R_{\mu\nu}(q)$, which is determined by the behavior of the solution near
$\chi=0$, cf. Eq.~(\ref{SR}). Physically, this is so since the current
disappears by decaying into a pair of massless fields, so like in the
vacuum, and the subsequent fate of these fields is irrelevant for the
calculation of the total decay rate (the imaginary part of $R_{\mu\nu}$).
This is similar to the calculation of the total cross--section for
$e^+e^-$ annihilation in lowest--order perturbation theory: this
cross--section is fully given by the $e^+e^-$ annihilation into a $q\bar
q$ pair, although the quark and the antiquark are not the actual final
states in the experiments. The `late stages' radiations or interactions
involving the quark and the antiquark, although essential for
hadronisation and the composition of the final state, do not affect the
total $e^+e^-$ cross--section.

\texttt{(ii)} At ultrarelativistic energies $k\gg K^3$, the medium
effects (as enhanced by the energy) become important already at the very
short radial distance $\chi_{AC}\sim {1}/{k^{1/3}}\ll 1/K$, meaning that
the current has no time to fully develop its partonic fluctuation before
getting absorbed. At very early times, such a fluctuation starts to
develop via diffusion, but this process is interrupted after the
relatively short period $\tilde t_s\sim k^{1/3}$, or
 \beq\label{ts}
 t_s\,\sim\,\frac{1}{T}\,
 \left(\frac{q}{T}\right)^{1/3} \
  \ll \ t_c\,\sim\, \frac{q}{Q^2} \,,
 \eeq
(which is still bigger than $1/T$ though), when the wave--packet has
diffused up to $\chi_{AC}$. Physically, this means that the growth of the
fluctuation is stopped at a transverse size
 \beq
 L_s \,\sim\,\sqrt{\frac{t_s}{q}}\,
 \sim\,\frac{1}{T}\,\left(\frac{T}{q}\right)^{1/3}\ \ll\ \frac{1}{Q}
 %\ \ll\ \frac{1}{T}
 \,,
 \eeq
which is much smaller than the natural size $\sim 1/Q$ for the same
fluctuation in the vacuum. Accordingly, the dynamics is very different
from the vacuum case, in the sense that the dissipation of the current
and the respective polarization tensor are now controlled by the medium.
This polarization tensor, which is now identical for space--like and
time--like currents, has been computed in Ref. \cite{Hatta:2007cs}, and
the result was used to deduce a partonic interpretation for the structure
of the plasma in its infinite momentum frame. In the next subsection, we
shall address this problem from a different perspective, by following the
time evolution of the Maxwell wave during its fall towards the black
hole. The insight that we shall gain in this way will allow us to
propose, in Sect. \ref{Concl}, a physical picture for the current
dissipation in the strongly coupled plasma.

\comment{ At this point it is convenient to introduce the {\em saturation
momentum} $Q_s(q,T)$, which played an important role in the analysis of
deep inelastic scattering in Ref. \cite{Hatta:2007cs}, and which can be
most generally defined as the current virtuality corresponding to the
region of transition between the low--energy and, respectively, the
high--energy regimes. The condition $qT^2\sim Q^3$ then implies
$Q_s(q,T)\sim (qT^2)^{1/3}$, and the above scales can be rewritten as
 \beq
 t_s\,\sim\,
 \frac{q}{Q^2_s(q)} \,,\qquad L_s \,\sim\,\frac{1}{Q_s(q)}\,.\eeq
This scale will appear again, in the analysis of the late--time behavior
and in the physical discussion in the next coming sections. }

\subsection{The fall of the wave into the black hole}
\label{Fall}

We now come to the most interesting physical situation, which refers to
the (relatively) late stages of the current evolution in the plasma and
the mechanism for current dissipation. In this situation, the dynamics is
controlled by the medium--dependent piece in the potential (the piece
$V_c\sim\chi^4$ in  Eq.~(\ref{VABC})), which yields the following
equation of motion in Schr\"odinger form
 \beq
 i\frac{\del  \psi}{\del\tilde t} \,=\,\left(-\frac{1}{2k}\,
 \frac{\del^2}{\del \chi^2}
 \,-\,\frac{k\chi^4}{32}\right)\psi\,. \label{SchLate}
 \eeq
As previously explained, for a time--like current this equation holds
both at moderate energies, $K^2\ll k\ll K^3$, and in the high--energy
limit $k\gg K^3$, but in ranges for $\chi$ which are different in the two
cases (cf. Eqs.~(\ref{Vlow})--(\ref{Vhigh})). For a space--like current,
it holds only for $k\gg K^3$ and $\chi\gg \chi_{AC}$.

We have not been able to find an exact solution to this equation, but we
shall construct a WKB approximation to it. To that aim, we use the
wave--packet representation in Eq.~(\ref{gauss}). The corresponding WKB
solution reads then
 \beq\psi(\tilde{t},\chi)&=&\int \rmd\varepsilon \,
 \rme^{-i\varepsilon (\tilde{t}-\tilde{t}_0)
 -\frac{\varepsilon^2}{2\sigma^2}}\,
 \frac{1}{p_\varepsilon(\chi)}\,\exp\Bigg\{i\int\limits_{\chi_0}^\chi
 \dif\chi'\,p_\varepsilon(\chi')\Bigg\}\,,\nn
 p_\varepsilon(\chi)&\,\equiv\,&\sqrt{2k(\varepsilon-V_C(\chi))}\,,
 \eeq
where we have kept only the outgoing wave and $\chi_0$ is either
$\chi_{BC}$, or $\chi_{AC}$, depending upon the physical regime under
consideration (and similarly for $\tilde{t}_0$). At this point, we notice
that the interesting values of $\tilde t$ are large enough for the
typical energy $\varepsilon\sim 1/\tilde t$ to be much smaller than the
potential $|V_C|\sim k\chi^4$. For instance, when $k\gg K^3$, we are
interested in $\tilde t\gtrsim \tilde t_s\sim k^{1/3}$ together with
$\chi\gg \chi_{AC}\sim{1}/{k^{1/3}}$. Then we can expand the square root
within $p_\varepsilon$ to linear order in $\varepsilon$ and perform the
ensuing Gaussian integration, to obtain
 \beq\label{psifall}
 \psi(\tilde{t},\chi)\,\simeq\,\frac{1}{p_0(\chi)}\,
 \,\exp\left\{i\int\limits_{\chi_0}^\chi\,
 \dif\chi'\,p_0(\chi')
 -\frac{\sigma^2}{2}\left[(\tilde{t}-\tilde{t}_0)
 -4\left(\frac{1}{\chi_0}-\frac{1}{\chi}\right)\right]^2\right\}.
 \eeq
We have also used here
 \beq
 \frac{\del p_\varepsilon}{\del \varepsilon}\bigg|_{\varepsilon=0}
 =\sqrt{\frac{k}{2|V_C|}}=\frac{4}{\chi^2} \ \ \ \Longrightarrow\ \
 \int\limits_{\chi_0}^\chi
 \dif\chi'\,\frac{\del p_\varepsilon}
 {\del \varepsilon}\bigg|_{\varepsilon=0}=\,
 4\left(\frac{1}{\chi_0}-\frac{1}{\chi}\right)
 \,.\eeq
The modulus $|\psi(\tilde{t},\chi)|$ tells us where the energy of the
wave--packet is located at time $\tilde{t}$. Clearly, with increasing
time, the peak of the energy distribution moves along a trajectory
 \beq\label{chifall}
  \chi(\tilde{t})\,=\,\frac{\chi_0}{1-
 \frac{\chi_0}{4}(\tilde{t}-\tilde{t}_0)}\ ,\eeq
which can be recognized as the trajectory of a classical particle with
mass $m=k$ and zero total energy moving in the potential $V_C(\chi)$.
This is so almost by construction (because the classical trajectory
defines the group velocity for the WKB solution), and can be also checked
by starting with the respective particle equation of motion, that is,
 \beq\label{part}
 k\,\frac{\dif^2\chi}{\dif \tilde{t}^2}\,=\,-\frac{\dif V_c}{\dif \chi}
 \,=\,k\,\frac{\chi^3}{8}\,,
 \eeq
whose zero--energy solution brings us back to Eq.~(\ref{chifall}), as it
should.

The approximation (\ref{chifall}) holds only so long as $\chi\ll 1$, but
it can be used to estimate the duration of the fall of the wave--packet
in the potential; namely, $\chi(\tilde{t})$ reaches the horizon at
$\chi=2$ when $\tilde{t}\sim \tilde{t}_h$, with
 \beq\label{tfall}
 \tilde{t}_h-\tilde{t}_0\,=\,\frac{4}{\chi_0}\,-\,2
 \,\simeq\,\frac{4}{\chi_0}\,.\eeq
Note that, parametrically, $\chi_0\sim 1/\tilde{t}_0$ for both the
intermediate--energy and the high--energy regimes. Thus,
$\tilde{t}_h-\tilde{t}_0\sim \tilde{t}_0$, so the total time
$\tilde{t}_h$ necessary for the wave--packet to propagate within $AdS_5$
from the boundary to the horizon is of order $\tilde{t}_0$. Essentially,
half of this time is used to (rather slowly) reach the ` point of no
return' at $\chi_0\ll 1$ (the point where the wave starts to feel the
attraction of the black hole) and the other half to travel along the
considerably larger distance from $\chi_0$ up to the horizon. Clearly,
for $\chi>\chi_0$ the motion of the wave--packet is accelerated by the
gravitational force. In fact, Eq.~(\ref{part}) implies that the group
velocity $\dot\chi\equiv {\dif\chi}/\dif \tilde{t}$ approaches the speed
of light (from below) when the wave gets close to the horizon:
 \beq\label{part1}
 {\dot\chi}\,=\,\frac{\chi^2}{4}\ \ \Longrightarrow\ \
 \dot\chi\simeq 1\quad\mbox{when}\quad\chi\simeq 2\,.
 \eeq
In reality, the particle cannot cross the horizon (in the laboratory
frame), but only asymptotically approach to it, so its velocity can never
become exactly one. But when $\chi\simeq 2$, the terms in the potential
which have been neglected in writing Eq.~(\ref{SchLate}) become important
and modify Eq.~(\ref{part1}). Let us also note here the `dual' version of
the above equation, which is particularly suggestive : using the
correspondence $\chi\sim T L$ and introducing the parton transverse
momentum $p_\perp$ via the uncertainty principle, i.e., $p_\perp\sim
1/L$, we arrive at
 \beq\label{force}
 \frac{\dif p_\perp}{\dif t}\,\sim \,-T^2\,.
 \eeq
That is, there is a transverse force acting on the colored partons in the
plasma, and this force is independent of the parton momentum and of order
$T^2$. This result will play an important role in the physical discussion
in Sect. \ref{Phys}.

\vspace*{.1cm} Let us conclude this section with a brief summary of the
previous results, in physical terms:

\vspace*{.2cm}

 \texttt{(i)} At {\em moderately high energies}, such that
$Q/T \ll q/Q\ll (Q/T)^2$, a time--like current creates SYM quanta in the
plasma at $t=0$, which then pass through three stages before disappearing
:

 \texttt{(a)} a relatively short period of {\em diffusion},
at  $t\lesssim t_c$ with $t_c\sim {q}/{Q^2}$, during which the current
gets replaced by a bunch of on--shell massless partons (only two such
partons being manifest in our calculation), with overall transverse size
$L\sim 1/Q$\,;

\texttt{(b)} a longer period of {\em free streaming}, from $t\sim t_c$ up
to $t_f \sim({1}/{T}) \sqrt{{q}/{Q}}$, during which the partonic system
propagates without feeling the plasma and expands in transverse space up
to a size $L\sim({1}/{T}) \sqrt{Q/q}\ll 1/T$, and

\texttt{(c)} an equally long period, during which the interactions with
the plasma cause an {\em accelerated expansion} in transverse space, up
to a maximal size $L\sim 1/T$; when this size is reached, the partons
move apart from each other at the speed of light ($v_\perp\simeq 1$) and
disappear in the plasma.

The lifetime of the current is controlled by the two last stages, and
thus is of order $t_f$, which is much larger than the coherence time
$t_c$.

\vspace*{.2cm} \texttt{(ii)} At {\em very high energies}, such that
$q/Q\gg (Q/T)^2$, a virtual current, either time--like or space--like,
goes through only two stages before dissipation: %\vspace*{.1cm}

\texttt{(a)} a relatively short period of {\em diffusion}, at $t\lesssim
t_s$ with $ t_s \sim {q}/{Q^2_s(q)}\ll t_c$, during which the partonic
fluctuation coming from the current grows up to a transverse size $L\sim
1/{Q_s(q)}$, which is however not sufficient for the partons to become
on--shell, and

\texttt{(b)} an equally short period of {\em accelerated expansion}, at
times $t > t_s$, which lasts until the partonic fluctuation reaches a
transverse size $L \sim 1/T$ and disappears in the plasma.

Both stages contribute on equal footing to the lifetime of the current,
so this lifetime is of order $t_s$.

Above, we have introduced the {\em saturation momentum} $Q_s(q)\sim
(qT^2)^{1/3}$, which is the natural transverse momentum scale to discuss
the high--energy scattering off the plasma \cite{Hatta:2007cs}. This is
defined as the `critical' virtuality for the transition between the
low--energy and the high--energy regimes: we have indeed $q/T\sim
(Q_s(q)/T)^3$. Note that, at high energy, $Q_s(q)$ replaces $Q$ as the
virtuality scale which determines all the physically relevant scales for
transverse size and time. (For instance, both $t_c$ and $t_f$ reduce to
$t_s$ after the replacement $Q\to Q_s(q)$.) Moreover, $Q_s(q)$ is also
the critical scale for the disappearance of the potential barrier in the
space--like case.

\section{Physical discussion}
\setcounter{equation}{0}\label{Concl}

This section will contain no new technical developments, but in spite of
that it could be viewed as the main section of this paper; indeed, here
is where, on the basis of the previous results, we shall develop our
physical picture. But before we do so, it is useful to make contact with
some previous approaches in the literature. This should make clear that
the physical picture that we shall later develop will also shed light on
some of the results of these previous approaches.

\subsection{Relation with previous approaches}
\label{Paral}

In this subsection, we shall show that our previous results are
consistent with, and shed new light on, previous studies in the
literature
\cite{Peeters:2006iu,Liu:2006nn,Chernicoff:2006hi,Caceres:2006ta,Avramis:2006em,Liu:2006he,Gubser:2008as,Herzog:2006gh,Gubser:2006bz},
concerning the propagation of `heavy quarks', or of quark--antiquark
`mesons', in the strongly coupled ${\mathcal N}=4$ SYM plasma.

To be specific, let us start with an example (we shall present some
others later on): Consider a time--like current at moderately high energy
($Q/T \ll q/Q\ll (Q/T)^2$) and focus on our estimate (\ref{Lg}) for $L_f$
--- the transverse size at which the partonic system created via the
decay of the current starts to lose energy in the plasma. After
introducing $v_z=q/\omega$ and $\gamma\equiv 1/\sqrt{1-v_z^2}=q/Q$, this
result can be rewritten as
 \beq\label{Lgg}
 L_f
 \,\sim\,\frac{1}{{\gamma}^{1/2} T}\,
\ \sim\ \frac{(1-v_z^2)^{1/4}}{T}\,,
 \eeq
which turns out to be the same as the parametric estimate for the `meson
screening length' $\ell_{\rm max}(v_z,T)$ computed in Refs.
\cite{Peeters:2006iu,Liu:2006nn,Chernicoff:2006hi}. The latter is the
maximal transverse size that a `heavy meson' (a bound system of a quark
and an antiquark) can have in the plasma, when it propagates along the
$z$ axis with a constant velocity $v_z<1$. So long as the transverse size
$\ell$ of the meson is smaller than $\ell_{\rm max}(v_z,T)$, the $q\bar
q$ pair is bound indeed and it moves freely through the plasma
--- that is, it moves at constant speed $v_z$ without the need for an
external force. Mathematically, this is described by a string connecting
the quark and the antiquark through $AdS_5$, which rigidly moves together
with its endpoints. For $\ell > \ell_{\rm max}(v_z,T)$, on the other
hand, this connecting--string solution ceases to exist and is replaced by
two disjoint strings trailing behind the quark and, respectively, the
antiquark, which extends in $AdS_5$ all the way up the horizon. This
trailing strings exerce a drag force on their fermionic sources, meaning
that energy is transferred from the quark and the antiquark to the black
hole \cite{Herzog:2006gh,Gubser:2006bz,Herzog:2006se,Caceres:2006dj}.

This meson dynamics is similar to that emerging from the present
calculation, except for the fact that, here, the transverse size of the
partonic pair is not a free parameter any longer, but is rather fixed by
the dynamics (and hence it depends upon time). Namely, over the time
interval $t_c < t <t_f$, which is large as compared to the formation time
$t_c\sim q/Q^2$ of the pair, the time--like current can be effectively
viewed as a `meson' made with a pair of on--shell, massless, partons
(Weyl fermions or adjoint scalars), which propagate together along the
$z$ axis with constant velocity $v_z$ and at the same time separate from
each other  in transverse directions with a (small) relative velocity
$2v_\perp$, with $v_\perp= \sqrt{1-v_z^2}\ll 1$. Hence, the transverse
size of the `meson' increases like $L=2v_\perp t$. When this size reaches
the critical value in Eq.~(\ref{Lg}) or (\ref{Lgg}) (this happens at a
time $t\sim t_f\simeq \sqrt{\gamma}/T$), the partons start to lose energy
to the plasma and then the `meson' disappears, so like in the
corresponding calculation in Refs.
\cite{Peeters:2006iu,Liu:2006nn,Chernicoff:2006hi}.

If this analogy is correct, then we should also find some similarity
between the trailing--string solutions in Refs.
\cite{Herzog:2006gh,Gubser:2006bz,Herzog:2006se,Caceres:2006dj} and the
dynamics of the current at {\em late} times $t\gg t_f$, where dissipation
is important. We will later show that this is indeed the case, but before
that let us emphasize some differences between our actual problem and
those in the previous literature:

$\bullet$ Our `partons' are massless fields from the Lagrangian of the
${\mathcal N}=4$ SYM theory. A pair of such fields is produced by the
current, at a distance $\chi\sim 1/K$ inside $AdS_5$ which is controlled
by the current virtuality. Physically, this means that this pair is
produced with a transverse size $L\sim 1/Q$. By contrast, in previous
studies, the `partons' are `heavy quark probes', that is, massive
fermions which do not belong to ${\mathcal N}=4$ SYM and whose
introduction requires an extension of the original AdS/CFT
correspondence: the dual partners of these fermions are open strings
ending on a D7 brane which extends along the radial direction of $AdS_5$
up to a distance $\chi$ inversely proportional to the fermion mass. This
makes it difficult to consider massless, or even light, quarks, since the
corresponding D7 brane would enter the black hole. Moreover, difficulties
appear even for heavy quarks in the high--energy limit ($v_z\to 1$),
since the latter appears not to commute with the limit in which the
ultraviolet cutoff is sent to infinity
\cite{Liu:2006ug,Argyres:2006vs,Liu:2006he,Argyres:2008eg}. This
particular problem does not appear in our formalism, where physics is
completely smooth in the limit $v_z\to 1$, even in crossing the
lightcone.

$\bullet$ The partonic pair produced by the decay of a time--like current
is, strictly speaking, not a `meson' (i.e., a bound state which would be
stable in the absence of the plasma), but rather a pair of jets, similar
to the $q\bar q$ jets created via $e^+e^-$ annihilation in perturbative
QCD. However, the relative velocity of these jets in transverse space is
much smaller than their common longitudinal velocity ($v_\perp=
\sqrt{1-v_z^2}\ll v_z$), so the fields remain close to each other for a
long time $\sim  \sqrt{\gamma}/T$; this may explain why their dynamics is
so similar to that of a meson. As we shall shortly argue, the partonic
fluctuation of a {\em space--like} (relativistic) current is even closer
to an actual meson.

$\bullet$ In our approach, there is no explicit string dual counterpart
of the `partons', or `jets' (so like the endpoints of the string on the
D7 brane in the `meson' problem). The partons appear only in the physical
interpretation on the gauge theory side, where they are supposed to be
generated via current fluctuations. Related to that, the 't Hooft
coupling $\lambda$, which in the other approaches is introduced by the
Nambu--Goto string action, is not present in our formalism, which in fact
corresponds to the strict strong--coupling limit $\lambda\to\infty$. This
feature sometimes complicates the comparison between our results and
previous approaches.

In spite of such differences, striking similarities persist (like the one
already discussed in relation with Eq.~(\ref{Lgg})), which support the
idea that the two types of problems (the `current' and the `meson') are
closely related. In what follows, we shall give some more examples in
that sense, which suggest that our wave--packet propagating through
$AdS_5$ --- the gravity dual of the plasma perturbation by a current ---
is in fact rather similar to the string solutions describing a meson, or
heavy quark, perturbation in the previous approaches.

Consider first a time--like current at moderate energies and not so late
times ($t< t_f$), corresponding to a relatively small meson. The dynamics
of the connecting--string solution on the `meson' side is characterized
by two important scales in $AdS_5$ (see, e.g.,
\cite{Chernicoff:2006hi,Liu:2006he}) : the position $\chi_{\rm
tip}(\ell)$ of the peak of the string (the maximal penetration of the
string in $AdS_5$) and the radial coordinate $\chi_v=2/\sqrt{\gamma}$
beyond which the string action would become imaginary. For a relatively
small meson $\ell \ll \ell_{\rm max}(v_z,T)$, one finds $\chi_{\rm
tip}(\ell)\sim T\ell$, which is precisely the relation between the
distance traveled in $AdS_5$ and the transverse size of the partonic
system that we advocated at the end of Sect. \ref{Coher}. This suggests
that one can identify $\chi_{\rm tip}(\ell)$ with the position of our
wave--packet in $AdS_5$. Furthermore, recalling that $\gamma=q/Q$, one
sees that $\chi_v\sim \sqrt{Q/q}$ is the same as our `point of no return'
$\chi_{BC}$, cf. Eqs.~(\ref{chis}) and (\ref{Vlow}) : beyond this point,
the wave cannot escape the attraction by the black hole.

Similar identifications can be made also for a {\em space--like} current
with moderate energy ($q\ll Q^3/T^2$): Then, $\chi_{\rm tip}(\ell)$ and
$\chi_v$ correspond to the first and, respectively, second `classical
turning points' for the potential barrier in Fig. \ref{PSlow}, i.e.,
$\chi_{AB}\sim T/Q$ and, respectively, $\chi_{BC}\sim \sqrt{Q/q}$. The
partonic system has now a fixed size $L\sim 1/Q$ (since the Maxwell wave
gets stuck at $\chi\lesssim \chi_{AB}$), and in that sense it looks even
closer to a real meson.

Returning to the case of a moderate--energy {\em time--like} current, let
us now consider the dynamics at later times, $t\gtrsim t_f$, where we
would like to make contact with the trailing--string solution in Refs.
\cite{Herzog:2006gh,Gubser:2006bz}. Notice first that the criterion for
the onset of dissipation, which determines the `critical' length
(\ref{Lgg}), is the same in our problem and in the `meson' problem: this
is the condition that the distance traveled in $AdS_5$ get close to the
`point of no return', which amounts to $\chi_g\sim\chi_{BC}$ (cf. the
discussion above Eq.~(\ref{tg})) in our context, and to $\chi_{\rm
tip}\sim \chi_v$ in the approach of Refs.
\cite{Peeters:2006iu,Liu:2006nn,Chernicoff:2006hi}. There is an
interesting difference though: whereas in the usual meson problem, this
change of regime is abrupt and characterized by a sharp critical value
$\ell_{\rm max}(v_z,T)$, in our approach it is much more
gradual\footnote{Similarly, for a space--like current, the transition
between the moderate--energy and the high--energy regimes, which happens
when $\chi_{AB}\sim \chi_{BC}$ (i.e., for $q\sim Q^3/T^2$), is smoothed
out by the tunnel effect \cite{Hatta:2007cs}.}, as it is driven by the
competition between the various terms in the potential. For
$\chi\gg\chi_{BC}$ (and hence $t\gg t_f$), our solution describes a
wave--packet falling into the black hole, along the trajectory in
Eq.~(\ref{chifall}). This should be compared to the trajectory of the
energy flow towards the horizon in the corresponding trailing--string
solution. The latter can be inferred from the results in Refs.
\cite{Herzog:2006gh,Gubser:2006bz} : the rate $\dif E/\dif t$ for energy
flow towards the horizon and the energy density $\dif E/\dif r$ per unit
length along the dragging string have been there computed as
(see\footnote{The authors of Ref. \cite{Herzog:2006gh} use the notation
$u$ for the radial coordinate in $AdS_5$, which is however not the same
as our variable $u$ in Eq.~(\ref{met1}); rather, their definition for
this variable is $u\equiv r/R^2$.}, e.g., Eqs.~(3.19a) and (3.20a) in
Ref. \cite{Herzog:2006gh})
 \beq\label{drag}
 \frac{\dif E}{\dif t}=\frac{\pi
 \sqrt{\lambda}\,T^2v_z^2}{2\sqrt{1-v^2_z}}
 \,,\qquad \frac{\dif E}{\dif r}=
 \frac{\sqrt{\lambda}}{2\pi\sqrt{1-v^2_z}R^2}\,. \eeq
(Recall that $\chi=2(r_0/r)$ with $r_0=\pi R^2 T$, and that we consider a
situation where $\chi\ll 2$, i.e., $r\gg r_0$.) These expressions involve
the 't Hooft coupling $\lambda$, which is inherent in the `heavy quark'
problem (it enters via the string tension, which provides the overall
normalization for the Nambu--Goto action, and also for the string
energy--momentum tensor) and seems to prevent any comparison to our
present results (which correspond to the limit $\lambda\to\infty$).
However, the coupling constant disappears in the ratio
 \beq\label{drdt}
 \frac{\dif r}{\dif t}\,\equiv\,-\frac{{\dif E}/{\dif t}}
 {{\dif E}/{\dif r}}\,=\,-\pi^2T^2R^2v^2_z\,, \eeq
which determines the trajectory of the energy flow in $AdS_5$, and is the
right result to be compared to our respective trajectory in Sect.
\ref{Fall}. And, indeed, Eq.~(\ref{drdt}), where for the present purposes
we can take $v_z\simeq 1$, is identical with our respective expression
(\ref{part1}), as it can be recognized after a change of variables.

In fact, the identification between our wave solution and the dragging
string can be made even more precise: the string world--sheet, as
expressed by the function $z(t,r)=v_zt-\zeta(r)$ (the string, which has a
shape $z=\zeta(r)$ in the variables $z$ and $r$, moves at constant speed
$v_z$, solidary with the heavy quark which is pulled by an external
force) is exactly the same as the surface of stationary phase for our
Maxwell wave solution $a(t,z,r)$, as computed in Ref.
\cite{Hatta:2007cs}. Specifically, in Ref. \cite{Hatta:2007cs} we have
studied the time--independent version of the above Eq.~(\ref{SchLate}),
but with a more general form for the potential, valid for all the values
of $\chi$ on the right of the `point of no return', up to $\chi=2$ (or
$r=r_0$). The plane--wave solution has been obtained in the WKB
approximation as (see Eq.~(D.4) in Appendix D of Ref.
\cite{Hatta:2007cs})
 \beq\label{trail}
 a(t,z,r)\,\approx\,\rme^{-i\omega t+iq z}\,\exp\left\{i
 \frac{q}{2\pi T}\left(\frac{1}{2}\ln \frac{r+r_0}{r-r_0}+ \arctan
 \left(\frac{r}{r_0}\right) \right)\right\}\,.\eeq
(For $r\gg r_0$ and after a change of variables, this expression reduces
indeed to the complex exponential part of Eq.~(\ref{psifall}).) The
condition of constant phase determines a hypersurface in $AdS_5$ which
can be written as $z-(\omega/q)t=\zeta(r)+const.$ where $\omega/q\simeq
1$ is the phase velocity and $\zeta(r)$ is the same function as the shape
of the trailing string, cf. Eq.~(3.17) of Ref. \cite{Herzog:2006gh}, or
Eq.~(10) of Ref. \cite{Gubser:2006bz}. Note the logarithmic singularity
in Eq.~(\ref{trail}) at $r\to r_0$ : this reflects the fact that, as
mentioned in Sect. \ref{Fall}, the wave--packet can never cross the
horizon, but only asymptotically approach to it, according to a law
$r-r_0\propto\exp(-4\pi T t)$. In the next subsection, we shall propose a
physical interpretation for this profile $\zeta(r)$ of the trailing
string.

At this point, we should also emphasize an important difference between
our present set--up and the `meson' problem in the previous literature:
In the latter, the transverse size $\ell$ of the meson and its
longitudinal velocity $v_z$ are independent parameters, so it is in
principle possible to choose a velocity $v_z$ arbitrarily close to one
(although in practice this meets with the problem of the UV cutoff, as
alluded to before
\cite{Liu:2006ug,Argyres:2006vs,Liu:2006he,Argyres:2008eg}). In our
approach, the `meson' is dynamically generated as a fluctuation of the
virtual current, which requires a formation time $t\sim t_c$. This
introduces an upper limit on the longitudinal momentum $q$ up to which a
nearly on--shell `meson' can form in the plasma: when the `meson'
lifetime $t_f \sim({1}/{T}) \sqrt{{q}/{Q}}$ becomes comparable to its
formation time $t_c$ --- this happens for $q\sim Q^3/T^2$ ---, the
partonic fluctuation melts in the plasma before having the time to become
on--shell. Hence, a `meson' can form only so far as $q\ll Q^3/T^2$, which
in turn introduces an upper limit on the longitudinal velocity $v_z$ of
the meson thus generated: $\gamma\ll (Q/T)^2$, or $1-v_z^2> (T/Q)^4$.
But, of course, there is no corresponding limit on the energy of the
current: when $q\gg Q^3/T^2$, the `point of no return' is reached after
the very short time $ t_s \sim {q}/{Q^2_s(q)}\ll t_c$, and then the
current disappears in the plasma before forming a meson. The late--time
dynamics which is responsible for this disappearance, as studied in Sect.
\ref{Fall}, is exactly the same for on--shell or off--shell partonic
fluctuations : it is always characterized by the group velocity
(\ref{part1}) for the falling wave--packet and by the surface of
stationary phase surface determined from Eq.~(\ref{trail}). This points
out towards the universality of the dissipation mechanism in a plasma at
strong coupling, which is the same for a colorful (heavy or massless)
quark as for a (sufficiently energetic) colorless meson or current,
although in the AdS/CFT calculation this dynamics may embrace different
mathematical descriptions (e.g., a trailing string, or a Maxwell--like
wave--packet).

In the next, final, subsection we shall try to elucidate the actual
physical mechanism which is responsible for this dissipation in the
strongly--coupled gauge theory.

\subsection{Parton branching at strong coupling}
\label{Phys}

In our physical discussion so far, we have privileged the interpretation
in which the time--like current first fluctuates (over a time of the
order of the coherence time $t_c\sim q/Q^2$) into a pair of massless
partons which subsequently undergo free motion, along straightline
trajectories. In the vacuum, this scenario would hold for ever, and
independently of the energy of the current, which only determines the
velocities of the particles in the pair. In the plasma, this would hold
only for moderately high energies, $q\ll Q^3/T^2$, and for values of $t$
smaller than the time $t_f\sim ({1}/{T}) \sqrt{{q}/{Q}}$ at which the
pair is large enough (in transverse direction) to feel the plasma. But
even for larger times $t>t_f$, the partons would still follow classical
trajectories, which now describe (in the dual theory) the fall of the
Maxwell wave into the black hole. At higher energies, $q\gg Q^3/T^2$,
this accelerated fall would show up right away after the early period of
diffusion.

Although consistent with our previous calculations and also quite
appealing due to its simplicity, this physical picture cannot be fully
right, for several reasons: First, at strong coupling, there is no reason
why the current should couple to two--parton final states alone, or why
these partons should undergo free motion, not even in the vacuum. Rather,
the two partons produced in the first, `electromagnetic', splitting can
in turn radiate other partons via strong, `color', interactions, and also
strongly couple to the fluctuations of the vacuum, thus giving rise to a
complicated, multi--partonic, final state. Second, the fact that the
partons appear to follow classical trajectories in the physical
(longitudinal and transverse) space contradicts the quantum nature of the
${\mathcal N}=4$ SYM theory. Although, in practice, we solve classical
equations of motion in the dual gravity theory, the results thus obtained
must somehow reflect the quantum nature of the original problem in gauge
theory, that should fully reveal itself at strong coupling. This is
already manifest in the dynamics at early stages, where our complex
exponential in Eqs.~(\ref{psivac})--(\ref{we}) can be recognized to
describe, via the correspondence $\chi\leftrightarrow \pi T L$, {\em
quantum} diffusion (i.e., quantum Brownian motion) in the physical
transverse space. There should be a corresponding quantum dynamics hidden
in the classical wave solution at later times $t> t_c$, and in what
follows we shall try to unveil this dynamics.

This seems {\em a priori} difficult because of the non--renormalization
property of the polarization tensor in the ${\mathcal N}=4$ SYM theory,
as discussed in Sect. \ref{Polar}. It is precisely this property which on
one hand makes the corresponding result to look so simple at strong
coupling (cf. Eq.~(\ref{Rvac})), but on the other hand forbids any direct
access to the detailed nature of the final state --- and hence to the
actual fate of the partons produced by the current. Here, we shall simply
try to reconstitute this fate from the dual wave solution via physical
considerations, notably, by using the uncertainty principle.

Note first that by using the uncertainty principle alone one can
correctly estimate the formation time for the partonic fluctuation, i.e.,
the coherence time $t_c$ alluded to above. Indeed, in the rest frame of
the current, where its 4--momentum reads $q^{\prime\mu}=(\omega',0,0,0)$
with $\omega'=Q$, the uncertainty principle requires a fluctuation time
$t'_c\gtrsim 1/Q$. This becomes $t_c=\gamma t'_c\sim q/Q^2$ after a
Lorentz boost to the frame in which the current is relativistic.

We now turn to later times $t > t_c$ (but lower than $t_f$ in the case of
the plasma), where the behavior in
Eqs.~(\ref{largetvac})--(\ref{groupvac}) applies, which in turn implies
that the partonic system expands in transverse space according to
Eq.~(\ref{Lfree}). Previously, we have interpreted these results as
describing the free motion of two classical particles, but now we shall
argue that they are also consistent with {\em quantum branching at strong
coupling}, a dynamics which is much more plausible given the
circonstances. In this interpretation, the expansion of the packet in
transverse space is the result of the degradation of the partons
transverse momentum via successive branching, in conformity with the
uncertainty principle.

Specifically, let us describe the branching process in terms of
generations --- a kind of mean field picture which should be reasonable
at large $N_c$ --- and start with a single parton (with longitudinal
momentum $q$ and virtuality $Q^2$) at $t_0=0$. We shall consider only a
$1\to 2$ splitting vertex, which is representative for the dynamics in a
gauge theory. Also, we shall assume that each of the two daughter
particles produced in a splitting takes a way, roughly, half of the
momentum and virtuality of the parent particle. This assumption is
natural at strong coupling, since there there is no need to look for
special corners in phase--space, so like collinear or soft emissions, to
enhance the probability for splitting. Thus, in the $n$th generation, as
obtained after $n\ge 1$ successive branchings, there will be $2^n$
particles, each of them carrying a longitudinal momentum $q_n\simeq
q/2^n$ and a virtuality\footnote{As before, the modulus on $Q$ and on
$Q_n$ is implicitly understood for time--like partons.} $Q_n\simeq
Q/2^n$. Since the virtuality was relatively small to start with ($q\gg
Q$), and it further decreases via the splitting, the particles in each
generation are nearly on--shell. At strong coupling, a splitting occurs
as fast as permitted by the uncertainty principle, so the lifetime of the
$(n-1)$-th generation --- the time $t_n-t_{n-1}$ necessary to go from the
$n-1$ to $n$ branching --- can be estimated from the energy imbalance at
the splitting vertex. This yields $t_n-t_{n-1}\sim q_n/Q_n^2\sim
2^n(q/Q^2)$, which rapidly grows with $n$. The time evolution of the
virtuality can therefore be estimated as
 \beq\label{dQdt}
 \frac{Q_n-Q_{n-1}}{t_n-t_{n-1}}\,
 \sim\,-\,\frac{Q}{q}\,Q_n^2
 \qquad\Longrightarrow\qquad \frac{\dif Q(t)}{\dif t}\,\simeq\,
 \,-\,\frac{Q^2(t)}{\gamma}\,,\eeq
where $\gamma=q/Q$ is the Lorentz factor for the original parton.
Introducing (by virtue of the uncertainty principle, once again) the
transverse size $L_n\sim 1/Q_n$ of the assemble of partons making up the
$n$th generation, and similarly $L(t)\sim 1/Q(t)$, we finally obtain
 \beq
 \frac{\dif L(t)}{\dif t}\,=\,\frac{C}{\gamma}\,,\eeq
with $C$ a number of order one whose precise value is not under control.
This is consistent, as anticipated, with the law (\ref{Lfree}) obtained
from the $AdS_5$ calculation via the correspondence $\chi\leftrightarrow
\pi T L$. Incidentally, Eq.~(\ref{dQdt}) implies $Q(t)\sim \gamma/t$,
valid for $t > t_c$.

In the vacuum of the conformal ${\mathcal N}=4$ SYM theory, this
successive branching would hold for ever, down to smaller and smaller
values of the longitudinal momentum fraction $z_n=q_n/q_{n-1}$ of the
produced partons. If, in order to mimic confinement, the theory is
supplemented with an infrared cutoff $\Lambda$ (say, in the form of a
cutoff $r_{\rm min}=\Lambda R^2$ --- implying a maximal value $\chi_{\rm
max}\propto 1/\Lambda$ --- on the radial distance in the dual string
theory), then the splitting will continue until the virtuality of the
softest produced partons will become of the order of this cutoff:
$Q_N\sim\Lambda$ . The total duration of the decay process will then by
controlled by the lifetime of the last generation, $t_N-t_{N-1}\sim
2^N(q/Q^2)\sim \gamma/\Lambda$, where we have used $2^N=Q/\Lambda$ and
$q/Q=\gamma$. The final partons produced in this process are relatively
numerous ($2^N=Q/\Lambda\gg 1$) and have small transverse momenta
$q_\perp\sim Q_N\sim \Lambda$, so they will be isotropically distributed
in transverse space, within a disk with area $\sim 1/\Lambda^2$ around
the longitudinal axis.

But in the case of a plasma, this vacuum--like branching continues only
up to a time $t_f \sim \sqrt{\gamma}/{T}$, when the partonic system has
expanded in transverse space up to a size $L_f\sim {1}/\sqrt{\gamma}{T}
\ll 1/T$. It is easy to understand this `critical' value in the current
scenario: at $t\sim t_f$, the softest partons in the cascade have a
typical virtuality $Q_N\sim 1/L_f\sim T \sqrt{\gamma}$ and a longitudinal
momentum $q_N=q/2^N\sim \gamma^{3/2} T$. (Here $N$ denotes the number of
generations up to a time $t_f$; hence, $2^N\sim QL_f$.) These values
satisfy $Q_N^3\sim q_NT^2$, which is precisely the condition for the
partons in this $N$th generation to start interacting with the plasma. In
Sect. \ref{Plasma}, we have associated this condition with a highly
energetic current, which starts to feel the plasma already at early
stages during its evolution, before having the time to decay into
on--shell partons. Here, we have so far considered a low energy current
($qT^2\ll Q^3$), which at the beginning decays in the same way as in the
vacuum, but whose evolution yields a system of partons for which the
condition $Q_N^3\sim q_NT^2$ is eventually satisfied (since the
successive branchings lead to a faster decrease in $Q_n^3$ as compared to
$q_n$). The physical meaning of this condition can perhaps be better
appreciated by noticing that the rate for the change in virtuality in
this $N$th generation (i.e., at time $t\sim t_f$) is of the order
 \beq\label{diffQ} \frac{\dif Q(t)}{\dif
 t}\,\bigg|_{t=t_f}\,\sim\,-T^2\,,
 \eeq
which looks like a natural order of magnitude for the transverse force
exerted by a plasma on colored partons. At a first sight, this might look
consistent with a `quasi--particle' picture for the strongly--coupled
plasma, in which the quasi--particles (thermal excitations) have typical
momenta $\sim T$ and are randomly distributed in space, with a typical
interparticle separation $\sim 1/T$. But then the random scattering
between the parton and these quasi--particles would increase the {\em
dispersion} $<p_\perp^2>-\bar p_\perp^2$ in its transverse momentum,
rather than uniformly decrease its {\em average} momentum $\bar
p_\perp\sim Q$. So, most likely, Eq.~(\ref{diffQ}) does not describe
thermal scattering, and its precise physical origin remains to be
clarified. One should also notice that the partons in this $N$th
generation have a relatively long lifetime $\sim \sqrt{\gamma}/{T}$, so
they can lose a significant part of their transverse momentum
$|Q_{N+1}-Q_N|\sim T \sqrt{\gamma}$ (before further decaying) even though
the respective dissipation rate is rather small, $|\dif Q/\dif t|\sim
T^2$.

For larger times $t> t_f$, the dynamics proceeds in the same way as it
would have proceeded at all times $t > 0$ if the current was sufficiently
energetic ($q\gtrsim Q^3/T^2$) to start with: namely, the partons keep
branching, but this branching is now accelerated by their momentum loss
towards the plasma, at an average rate $\sim T^2$. Indeed, as already
noticed in relation with Eq.~(\ref{force}), the AdS/CFT result
(\ref{part1}) for the fall of the Maxwell wave into the back hole is
tantamount (after the identification $\chi\longleftrightarrow TL\sim
T/Q(t)$) to a decelerating transverse force
 \beq
 \frac{\dif Q(t)}{\dif t}\,\sim \,-T^2
 \,,\eeq
acting on the partons. Because of this force, the partons can now
dissociate much faster than in the vacuum : the lifetime $\Delta t_n\sim
q_n/Q_n^2$ of a parton from the $n$th generation (with $n
> N$) is determined by the condition that, during this time, the parton
be able to lose a transverse momentum of order $Q_n$, at a constant rate
$\sim T^2$. This condition determines the parton virtuality as $Q_n\sim
(q_nT^2)^{1/3}$, which is recognized as the `saturation momentum'
introduced in Sect. \ref{Plasma}
--- here evaluated for a parton with momentum $q_n\sim q/2^n$. Note that
this value for $Q_n$ is considerably larger than $Q/2^n$ (the
corresponding value in the vacuum after the same number of generations),
showing that the partons which are involved in this {\it medium--induced
branching} are highly off--shell.

It is also interesting to evaluate the rate $\dif q/\dif t$ for
longitudinal momentum loss in the plasma, i.e., the longitudinal force
acting on the parton. We can write
 \beq\label{dqdt}
 \frac{q_n-q_{n-1}}{t_n-t_{n-1}}\,
 \sim\,-\,\frac{q_n}{q_n/Q_n^2}\,\sim\,-\,Q_n^2
 \qquad\Longrightarrow\qquad \frac{\dif q(t)}{\dif t}\,\simeq\,
 \,-\,(qT^2)^{2/3}\,,\eeq
which is recognized as the `dynamical' version of the drag force computed
in Refs. \cite{Herzog:2006gh,Gubser:2006bz}. Namely, in those papers, one
has considered a heavy quark moving through the plasma at a constant
speed $v_z$ and found that, in order to compensate for its energy loss,
one needs to drag this quark with a constant force $\dif p/\dif t =
(1/v_z)(\dif E/\dif t)$ that can be read off the first equation
(\ref{drag}). Therefore\footnote{We take $v_z\simeq 1$, as appropriate
for comparing to our present calculations.} $\dif p/\dif t\propto \gamma
T^2$, where $\gamma =1/\sqrt{1-v^2_z}$ is time--independent in the
context of Refs. \cite{Herzog:2006gh,Gubser:2006bz}. Our force
(\ref{dqdt}) can be rewritten in a formally similar way by introducing
the Lorentz factor for the (virtual) parton as $\gamma(t)=q(t)/Q(t)$ ---
which however is now time--dependent, because of the uncompensated energy
loss --- and recalling that, for this parton, $Q(t)=Q_s(q(t))\sim
(qT^2)^{1/3}$, so that we indeed have $(qT^2)^{2/3}\sim \gamma(t) T^2$.

Eq.~(\ref{dqdt}) tells us that a parton (or current) will totally lose
its energy over a time interval $t_{\rm loss}\sim (1/T)(q_N/T)^{1/3}$,
with $q_N$ the value of its longitudinal momentum at the time where that
parton (or current) has started to feel the plasma. For an ${\mathcal
R}$--current with moderately high energy ($qT^2\ll Q^3$), we have seen
that $q_N \sim \gamma^{3/2}T$, which in turn implies $t_{\rm loss}\sim
\sqrt{\gamma} /T$, in agreement with the respective AdS/CFT result in
Eq.~(\ref{tg}). But for a very energetic current ($qT^2\gg Q^3$), and
also for a colored parton with any energy, $q_N$ is the same as its
original momentum at $t=0$, and then we find $t_{\rm loss}\sim
(1/T)(q/T)^{1/3}$, in agreement with the corresponding result in
Eq.~(\ref{ts}) and also with a very recent result in Ref.
\cite{Gubser:2008as}.

\FIGURE[t]{
\includegraphics[height=9.cm]{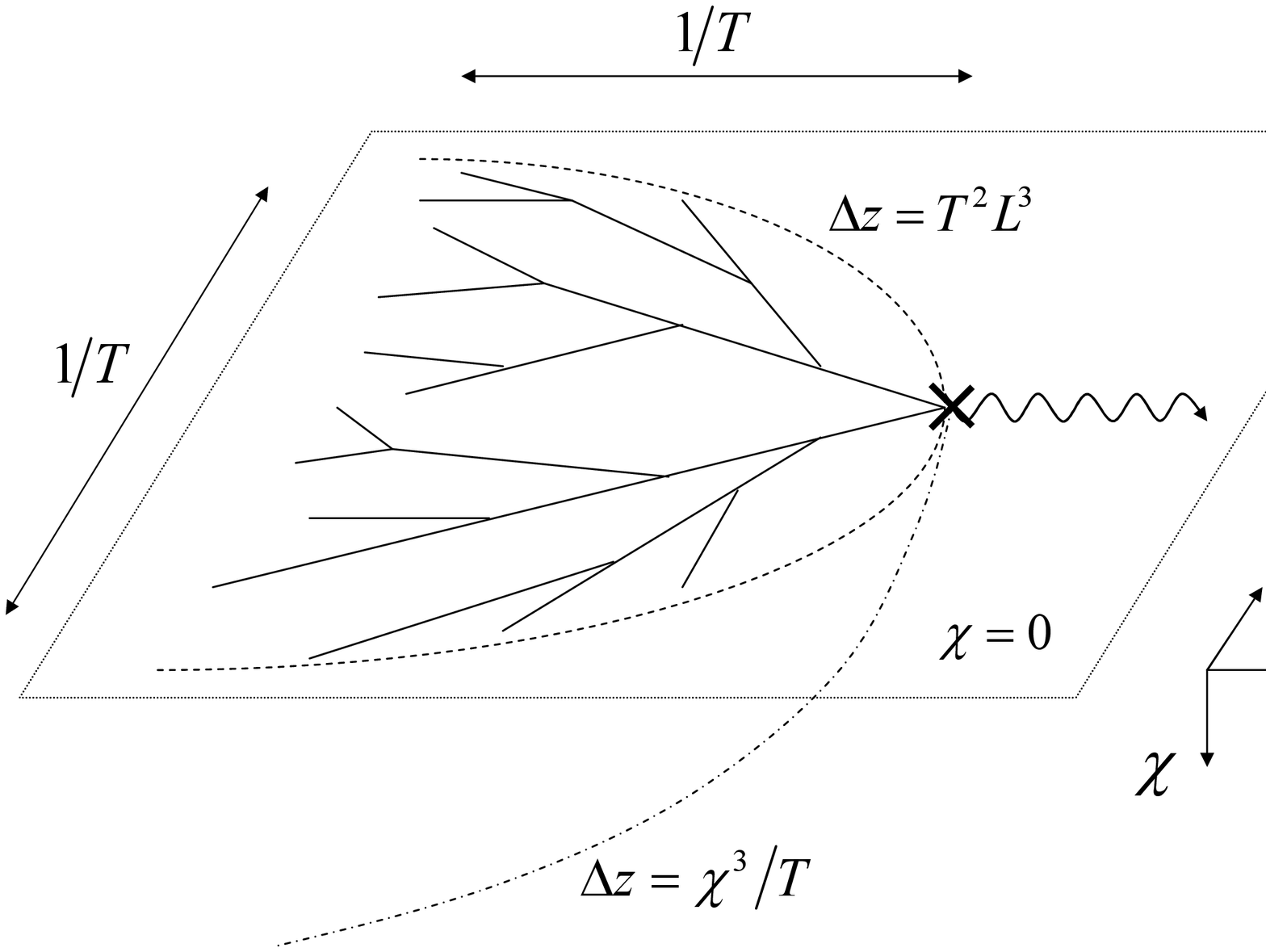}
\caption{\sl The parton cascade generated by the current via
medium--induced branching in the physical Minkowski space (represented
here as the boundary of $AdS_5$ at $\chi=0$) and the trailing string
attached to the leading particle (represented for $\chi\ll 1$). The
latter is `dual' to the enveloping curve of the former. \label{Figtrail}
}}

We shall conclude this discussion with a physical interpretation for the
trailing string solution originally obtained in Refs.
\cite{Herzog:2006gh,Gubser:2006bz} (and which, we recall, also emerges
from our Maxwell wave solution, cf. Eq.~(\ref{trail})). Namely, we shall
argue that the function $\zeta(r)$ which defines the shape of the string
is `dual', via the identification $r/R^2\leftrightarrow 1/L$, to the
enveloping curve of the parton distribution created via branching in the
medium. This curve can be best visualized if one considers a stationary
situation similar to that in Refs. \cite{Herzog:2006gh,Gubser:2006bz},
namely the situation in which some external force is continuously giving
energy to the system, in such a way that the leading particle propagates
at constant speed $v_z\lesssim 1$. The `leading particle' is either our
${\mathcal R}$--current in the high--energy regime at $qT^2\gg Q^3$, or
some colored parton like the heavy quark considered in
\cite{Herzog:2006gh,Gubser:2006bz}. This particle creates a parton
cascade via successive branchings and the average (gross) properties of
this cascade are independent of time. The parton distribution being
roughly isotropic in transverse space, as argued before, we can focus on
a particular plane, say, $(x,z)$  (see Fig. \ref{Figtrail}). The
enveloping curve of the distribution is then the function $\Delta z(L)$
which relates the longitudinal separation $\Delta z$ between a parton in
this plane and the leading particle to the transverse width $L$ of the
cascade at the position of that parton. To construct this function, it is
convenient to return to the description of the parton cascade in terms of
generations; this implies
 \beq
 \Delta z(L_n)\,\simeq\,\sum_{j=1}^n \,(1-v_j)\Delta t_j\,,\eeq
where $L_n\simeq 1/Q_n$ and where $Q_j$, $\Delta t_j$, and $v_j$ denote,
respectively, the typical virtuality, lifetime, and velocity of a parton
from the $j$th generation, with $j\le n$. According to the previous
discussion, we have $Q_j\sim (q_jT^2)^{1/3}$, $\Delta t_j\sim q_j/Q_j^2$,
and
 \beq
 v_j\,\simeq\,\frac{q_j}{\omega_j}\,=\,\frac{q_j}{\sqrt{Q_j^2+q_j^2}}
 \,\simeq\,1-\frac{Q_j^2}{2q_j^2}\,.\eeq
Therefore,
 \beq
 \Delta z(L_n)\,\sim\,\sum_{j=1}^n \,\frac{1}{q_j}
 \,\simeq\,\int_0^n\dif j \,\frac{1}{q(j)}
 \,\simeq\,\int_0^{L_n}\frac{\dif L}{L} \,T^2L^3
 \,\sim\,T^2L_n^3\,,\eeq
where we have used $\dif j\simeq \dif L_j/L_j$ (since $\dif (\ln
L_j)\simeq -(1/3)\dif (\ln q_j)\sim \dif j\,$; we recall that
$q^j=q/2^j$). We thus have found that $\Delta z(L)\sim T^2L^3$. After
replacing $L\to R^2/r$ within this result (in order to compare with the
shape $\zeta(r)$ of the trailing string), we finally deduce $\Delta z(r)
\sim T^2(R^6/r^3)$. As anticipated, this is parametrically the same as
the expansion of the function $\zeta(r)$ for $r\gg r_0$ (where our
present calculation is supposed to apply).

By inverting the function above, we find that $L$ grows with $\Delta z$
as $L\sim (\Delta z/T^2)^{1/3}$. This is correct so long as $L$ remains
smaller than $1/T$, and hence for values of $\Delta z$ which are
themselves smaller than $1/T$. But the transverse width of the partonic
distribution is roughly limited\footnote{Actually, as pointed out in
Refs. \cite{Gubser:2007xz,Chesler:2007an,Chesler:2007sv}, the energy and
momentum transfer from the leading particle can also induce longer range
($L \gg 1/T$) perturbations into the plasma, so like sound waves or Mach
cones. But in order to see such perturbations, one has to take into
account the feedback of the trailing string, or of the Maxwell wave, on
the $AdS_5$--Schwarzschild geometry, something that goes beyond the
purpose of this work.} to $L\lesssim 1/T$, since partons with momenta
$Q\lesssim T$ disappear in the thermal bath. This argument suggests that
at larger distances $\Delta z\gtrsim 1/T$ the parton distribution should
approach a cylinder with diameter $L\sim 1/T$ (see Fig. \ref{Figtrail}).
This is again similar to the corresponding behavior of the trailing
string, which for $\zeta\gtrsim 1/T$ approaches asymptotically the
horizon at $r= r_0$ \cite{Herzog:2006gh,Gubser:2006bz}.

This `duality' between the shape of the trailing string and the
enveloping curve of the parton distribution implies a similar duality
between our Maxwell wave in $AdS_5$ and the wave which would describe the
physical parton distribution in Minkowski space. Namely, the latter can
be obtained by replacing $r\to 2R^2/L$ in Eq.~(\ref{trail}), thus
yielding
 \beq\label{duala}
 a(t,z,L)&\,\approx\,&\rme^{-i\omega t+iq z +iq\Delta z(L)}\,,\nn
 \Delta z(L)&\,\equiv\,&
 \frac{L_0}{4}\left[\frac{1}{2}\ln \frac{L_0+L}{L_0-L}+ \arctan
 \left(\frac{L_0}{L}\right)\right],\qquad L_0\equiv\frac{2}{\pi T}\,.
 \eeq
The phase of this wave is stationary along a paraboloid in Minkowski
space which represents the enveloping surface of the parton distribution.
The fact that this distribution can be characterized by its enveloping
curve alone (i.e., that is does not depend separately upon the transverse
coordinates $x$ and $y$, with $x,y\le L$, but only upon the points on a
circle with radius $L$) can be simply understood in physical terms: all
the points in a transverse cross--section of this paraboloid at a fixed
value $\Delta z\equiv z-(\omega/q)t$ correspond to partons which belong
to a same generation. Such partons move in phase with each other, as they
have the same velocity. Accordingly, the phase of the wave must be the
same for all such points, as indeed happens for the wave in
Eq.~(\ref{duala}).

%\newpage

\section*{Acknowledgments}

We would like to thank I. Bena for useful discussions. Y.H. acknowledges
the hospitality of Institut de Physique Th\'eorique de Saclay, where he
started working on this particular problem. Y.H. and E.I. would like to
thank the organizers of the Yukawa International Program for
Quark--Hadron Sciences (YIPQS) {\em New Frontiers in QCD 2008}, hold at
Yukawa Institute for Theoretical Physics, for hospitality and support
during the late stages of this work. On this occasion they have benefited
from many useful discussions and insightful remarks, in particular, from
L. McLerran, B. M\"uller, H. Nastase, D. Teaney, and R. Venugopalan.
Special thanks to D. Teaney for explaining his work to us in detail. The
work of A.H.~M. is supported in part by the US Department of Energy. The
work of E.~I. in supported in part by Agence Nationale de la Recherche
via the programme ANR-06-BLAN-0285-01.

 \appendix
\section{Wave--packet evolution in the vacuum}
\setcounter{equation}{0}

In this Appendix we shall consider the initial value problem for the
Schr\"odinger--like equation (\ref{Lvac}) in a slightly more systematic
way, by using the {\em Ansatz} (\ref{gauss}) in order to search for
appropriate solutions. We recall that we are interested in solutions
$\psi(\tilde{t},\chi)$ which for $\tilde{t}$ are localized near $\chi=0$.
The precise form of the initial condition is however irrelevant, and in
what follows it will be more convenient (since mathematically simpler)
not to stick to a particular initial condition, but rather identify a
particular solution and then verify that this solution satisfies indeed
the condition of localization at early times ($\tilde{t}\to 0$).

We consider the time--like case, i.e., we take the minus sign in front of
$K^2$ in Eq.~(\ref{Lvac}), since this is the most interesting case. After
inserting the {\em Ansatz} (\ref{gauss}) into Eq.~(\ref{Lvac}), one can
recognize the Bessel equation for $\Psi(\varepsilon,\chi)$, up to some
trivial changes of function and variable. An interesting particular
solution is
 \beq\label{wpvac} \psi(\tilde{t},\chi)=\int \rmd\varepsilon \,
\rme^{-i\varepsilon \tilde{t}-\frac{\varepsilon^2}{2\sigma^2}}
\sqrt{\chi}\, \mathrm{J}_0(\chi\sqrt{K^2+2k\varepsilon}\,)\,,
 \eeq
where the reason for choosing the particular Bessel function
$\mathrm{J}_0$ (rather than the most general combination
$c_1\mathrm{J}_0+c_2\mathrm{N}_0$) should become clear in a moment. We
shall estimate the above integral separately for small and, respectively,
large values for the external coordinates $\tilde{t}$ and $\chi$
--- the change in behavior occurring at $\tilde t\sim \tilde t_c\simeq
2k/K^2$ (the coherence time) and for $\chi\sim\chi_c\simeq 1/2K$ (the
position of the wave--packet at $\tilde t= \tilde t_c$). In this
analysis, one should keep in mind that the width $\sigma$ of the packet
obeys to Eq.~(\ref{gau}).

Specifically, for small times $\tilde t\ll \tilde t_c$ (with $\tilde
t\gtrsim 1/\sigma$ though) and for $\chi\ll \chi_c$, the integral is
dominated by energies $\varepsilon$ such that
$K^2/2k\ll\varepsilon\ll\sigma$, as we shall shortly check. Then one can
neglect the Gaussian factor inside the integrand and also the $K^2$ term
inside the argument of the Bessel function. The resulting integral can be
performed exactly (see formula 6.631-6 in Ref. \cite{Gradshteyn94}), and
reproduces the corresponding solution found in Sect. \ref{Coher} (cf.
Eq.~(\ref{psivac})) :
 \beq\label{inteps} \psi\,\simeq\, \int_0^\infty \rmd
  \varepsilon \,\rme^{-i\varepsilon \tilde{t}}\, \sqrt{\chi}\,
  \mathrm{J}_0(\chi\sqrt{2k\varepsilon})
  =
  %2\sqrt{\chi}\int_0^\infty \rmd y \,y\,
  %\rme^{-iy^2\tilde{t}}J_0(\chi\sqrt{2k} y)=
  \frac{-i\sqrt{\chi}}{\tilde{t}}\,
  \rme^{i\frac{k\chi^2}{2\tilde{t}}}\,. \eeq
In fact, it was precisely this possibility, to exactly perform the above
integration, which motivated our choice for the Bessel function
$\mathrm{J}_0$ in Eq.~(\ref{wpvac}).

As discussed in Sect. \ref{Coher}, this solution implies that the energy
diffuses along a trajectory $\chi\sim\sqrt{\tilde{t}/k}$. For values of
$\chi$ along, or near, this trajectory, we have
$\chi\sqrt{2k\varepsilon}\sim \sqrt{\tilde{t}\varepsilon}$, and then the
integral in Eq.~(\ref{inteps}) is controlled by $\varepsilon\sim
1/\tilde{t}$ (since this is the only scale inside the integrand). Hence,
for $\tilde t\ll \tilde t_c$, we have $\varepsilon\gg K^2/2k$, as
anticipated. Since, moreover, the time variable is restricted to $\tilde
t \gtrsim 1/\sigma$, the condition that $\varepsilon\ll\sigma$ is
satisfied as well.

In the other interesting case, namely for relatively large times and
radial coordinates, $\tilde t\gg \tilde t_c$ and $\chi\gg \chi_c$, the
integral in Eq.~(\ref{wpvac}) is dominated by relatively small values of
$\varepsilon$, such that $2k\varepsilon\ll K^2$. Moreover, in this regime
we have $K\chi\gg 1$, so one can use the asymptotic expression for the
Bessel function, to obtain
\beq \psi \sim \int \rmd\varepsilon \, \rme^{-i\varepsilon
 \tilde{t}-\frac{\varepsilon^2}{2\sigma^2}+i\chi\sqrt{K^2+2k\varepsilon}}\,,
 \eeq
where we have kept only the outgoing--wave component of the solution.
(The other component would yield a solution propagating from the bulk
towards the boundary.) The dominant contribution can now be obtained by
expanding the square root in the exponent to linear order in
$\varepsilon$ and then performing a Gaussian integral :
\beq\label{expeps} \int \rmd\varepsilon \, \rme^{-i\varepsilon
\tilde{t}-\frac{\varepsilon^2}{2\sigma^2}+iK\chi+i\frac{k\chi}{K}\varepsilon
}\, \sim\,
\rme^{iK\chi-\frac{\sigma^2}{2}\left(\tilde{t}-\frac{k}{K}\chi\right)^2}\,.
\eeq
The strength $|\psi|$ of the wave, which shows where the energy is
located, has a peak at $\tilde{t}=({k}/{K})\chi$. Hence, this solution
represents a wave--packet with propagates in $AdS_5$ with constant group
velocity :
 \beq \chi=v_g\tilde{t}, \qquad
 v_g=\frac{K}{k}=\frac{1}{\gamma}=\sqrt{1-v^2_z}\,, \label{free1}
 \eeq
in agreement with the simpler analysis in Sect. \ref{Coher}.

\section{Classical particle falling in AdS}
\setcounter{equation}{0}

In this appendix we consider the motion of a classical pointlike particle
falling down into the $AdS_5$ black hole and compare its motion to the
propagation of a time--like current, as previously studied in Sect.
\ref{Plasma}. (See Ref. \cite{Sin:2004yx} for a similar study.)

A massless particle propagates along a null geodesics
  \beq 0=\rmd
s^2\propto -\left(1-\frac{\chi^4}{\chi_H^4}\right)\rmd t^2+\rmd x^2+\rmd
y^2+\rmd z^2+\frac{\rmd\chi^2}{(2\pi T)^2\left(1-\frac{
\chi^4}{\chi_H^4}\right)}\,, \label{app}
  \eeq
where $\chi_H=2$. For a particle with velocity (the variables $q$ and
$\omega$ are introduced below to facilitate the comparison with the
discussion in Sect. \ref{Plasma})
 \beq v=\frac{\rmd z}{\rmd t}\equiv \frac{q}{\omega}\,,
 \eeq
in the $z$--direction, (\ref{app}) reads
 \beq
\left(\frac{\rmd\chi}{\rmd\tilde{t}}\right)^2=
\left(1-\frac{\chi^4}{\chi_H^4}\right)\left(1-\frac{\chi^4}{\chi_H^4}-v^2
\right)\,. \label{app2} \eeq

In the static case, $v=0$, one can integrate explicitly
 \beq
\tilde{t}=\tanh^{-1}\frac{\chi}{\chi_H}+\tan^{-1}\frac{\chi}{\chi_H}\,.
 \eeq
The particle reaches near the horizon $\chi\sim \chi_H$ in time $t \sim
1/T$. This is indeed similar to the evolution of a  time--like current
with $q=0$.

When $q\neq 0$, (\ref{app2}) can be rewritten as
 \beq\label{geodesic}
\left(\frac{\rmd\chi}{\rmd\tilde{t}}\right)^2=
\left(1-\frac{\chi^4}{\chi_H^4}\right)
\left(\frac{|Q^2|}{\omega^2}-\frac{\chi^4}{\chi_H^4}\right)\,.
 \eeq
When $1\gg Q/\omega\gg \chi^2/\chi_H^2$, or
 \beq \chi \ll
2\sqrt{\frac{Q}{q}}\sim \chi_v
 \eeq
one easily finds
 \beq \frac{\rmd\chi}{\rmd\tilde{t}}\simeq \frac{Q}{\omega}\,,
\label{hen}
 \eeq
which is equivalent to Eq.~(\ref{groupvac}). This is valid until a time
 \beq t \lesssim \frac{1}{T}\sqrt{\frac{\omega}{Q}} \sim t_f\,,
 \eeq
when the `point of no return' $\chi \sim \chi_v$ is reached. The above
estimates for $t_f$ and $\chi_v$ are indeed consistent with the
corresponding ones derived from the current dynamics in Sect.
\ref{Plasma} (where $\chi_v$ was rather denoted as $\chi_{BC}$).
Therefore, the intermediate stage of the evolution of a time--like,
moderate energy ($q\ll Q^3/T^2$), `photon' looks indeed like a massless
pointlike particle falling into the bulk. On the other hand, the initial,
diffusive regime of the photon evolution has no classical analog.
Moreover, a pointlike particle stalls in the $\chi$ direction at
$\chi\sim \chi_f$, as clear by inspection of Eq.~(\ref{geodesic}),
whereas a photon (string) keeps on falling further into the bulk as
discussed in the main text.

%\bibliographystyle{utcaps}
%\bibliography{jetref}

\providecommand{\href}[2]{#2}\begingroup\raggedright\endgroup

\end{document}